\documentclass[twocolumn]{aastex7}
\usepackage{times,amsmath}
\usepackage[T1]{fontenc}

\hypersetup{pdfauthor={Kamlesh Rajpurohit \& Ewan O'Sullivan},
            pdftitle={A MeerKAT view of the Neutral Atomic Gas in Stephan's Quintet},
            pdfkeywords={Galaxy collisions; Galaxy interactions; Galaxy groups; Hickson compact group; HI line emission; Intergalactic medium; Intergalactic filaments; Shocks},
            bookmarksnumbered=true}
\pdfoutput=1

\usepackage[utf8]{inputenc}
\usepackage{graphicx}
\usepackage{amssymb}
\usepackage{amsmath}
\usepackage{multirow}
\usepackage{textcomp}
\usepackage{gensymb}
\usepackage{enumitem}  
\usepackage{hyperref}
\usepackage{natbib}
\usepackage{comment}
\usepackage{systeme}
\usepackage[Symbol]{upgreek}
\usepackage{xcolor}
\usepackage{xspace}
\definecolor{xlinkcolor}{cmyk}{1,1,0,0}

\newcommand{\arcm}{\hbox{$^\prime$}}

\newcommand{\chandra}{\emph{Chandra}}

\newcommand{\hst}{\emph{HST}}
\newcommand{\jwst}{\emph{JWST}}
\newcommand{\arcs}{\mbox{\arcm\arcm}}

\newcommand{\Msol}{\ensuremath{\mathrm{~M_{\odot}}}}

\newcommand{\s}{\ensuremath{\mbox{~s}}}
\newcommand{\ps}{\ensuremath{\s^{-1}}}
\newcommand{\cm}{\ensuremath{\mbox{~cm}}}

\newcommand{\pcmcu}{\ensuremath{\cm^{-3}}}

\newcommand{\km}{\ensuremath{\mbox{~km}}}

\newcommand{\kmps}{\ensuremath{\km \ps}}

\newcommand{\gtsim}{\,\rlap{\raise 0.5ex\hbox{$>$}}{\lower 1.0ex\hbox{$\sim$}}\,} 

\newcommand{\Hi}{H\textsc{i}}

\newcommand{\hi}{\textsc{H$\,$i}\xspace}

\newcommand{\SQ}{SQ}
\newcommand{\kms}{$\rm km\,s^{-1}$}

\shorttitle{MeerKAT view of H\textsc{i} in SQ}

\begin{document}

\title{A MeerKAT view of the Neutral Atomic Gas in Stephan's Quintet}

\correspondingauthor{Kamlesh Rajpurohit, Ewan O'Sullivan}
\email{kamlesh.rajpurohit@cfa.harvard.edu, eosullivan@cfa.harvard.edu }
\author[0000-0001-7509-2972]{K. Rajpurohit} 
\altaffiliation{These authors contributed equally to this work}
\affil{Center for Astrophysics $|$ Harvard \& Smithsonian, 60 Garden Street, Cambridge, MA 02138, USA}
\email{kamlesh.rajpurohit@cfa.harvard.edu}

\author[0000-0002-5671-6900]{E. O'Sullivan}
\altaffiliation{These authors contributed equally to this work}
\affil{Center for Astrophysics $|$ Harvard \& Smithsonian, 60 Garden Street, Cambridge, MA 02138, USA}
\email{eosullivan@cfa.harvard.edu}

\author[0000-0002-4962-0740]{G. Schellenberger}
\affil{Center for Astrophysics $|$ Harvard \& Smithsonian, 60 Garden Street, Cambridge, MA 02138, USA}
\email{gerrit.schellenberger@cfa.harvard.edu}
\author[0009-0007-0318-2814]{J. M. Vrtilek}
\affil{Center for Astrophysics $|$ Harvard \& Smithsonian, 60 Garden Street, Cambridge, MA 02138, USA}
\email{jvrtilek@cfa.harvard.edu} 
\author{L. P. David}
\affil{Center for Astrophysics $|$ Harvard \& Smithsonian, 60 Garden Street, Cambridge, MA 02138, USA}
\email{ldavid@head.cfa.harvard.edu}
\author[0000-0002-1634-9886]{S. Giacintucci}
\affil{Naval Research Laboratory, 4555 Overlook Avenue Southwest, Code 7213, Washington, DC 20375, USA}
\email{simona.giacintucci.civ@us.navy.mil}
\author[0000-0002-7607-8766]{P. N. Appleton}
\affil{Caltech/IPAC, MC 314-6, 1200 E. California Blvd., Pasadena, CA 91125, USA}
\email{pnapple@caltech.edu}
\author[0000-0002-1588-6700]{C. K. Xu}
\affil{Chinese Academy of Sciences South America Center for astronomy, National Astronomical Observatories, CAS, Beijing 100101, People's Republic of China}
\affil{National astronomical Observatories, Chinese Academy of Sciences, 20A Datun Road, Chaoyang District, Beijing 100101, People's Republic of China}
\email{coxu@ipac.caltech.edu}
\author[0000-0003-0202-0534]{C. Cheng}
\affil{Chinese Academy of Sciences South America Center for astronomy, National Astronomical Observatories, CAS, Beijing 100101, People's Republic of China}
\affil{National astronomical Observatories, Chinese Academy of Sciences, 20A Datun Road, Chaoyang District, Beijing 100101, People's Republic of China}
\affil{CAS Key Laboratory of Optical Astronomy, National Astronomical Observatories, Chinese Academy of Sciences, Beijing 100101, People's Republic of China}
\email{chengcheng@nao.cas.cn}
\author[0000-0003-1078-2539]{T. Deb}
\affiliation{Center for Astrophysics $|$ Harvard \& Smithsonian, 60 Garden Street, Cambridge, MA 02138, USA}
\email{tirna.deb@cfa.harvard.edu}
%

\begin{abstract} 
We present new MeerKAT 21cm spectral line observations of the neutral hydrogen gas in the compact galaxy group Stephan's Quintet (HCG~92). These data provide a significantly improved view of the atomic gas distribution and kinematics in the group. New features include the first detections of \Hi\ associated with member galaxies NGC~7319 and NGC~7320C, the identification of an additional high-velocity \Hi\ component associated with SQ-A, and the detection of additional \Hi\ at low velocities filling much of the area of the NGC~7318B disk. We also find \Hi\ in the previously detected gas bridge linking NGC~7319 and NGC~7318B, and a new northern bridge linking NGC~7319 to the SQ-A star-formation region. We detect \Hi\ with a wide range of velocities along the line of sight through the northern half of the famous shock ridge, including in the 6200-6500\kmps\ velocity range occupied by shocked H$\alpha$ emission. We examine the morphology and velocity structure of the \Hi\ and consider the origins of different components, finding some evidence that while the gas associated with NGC~7318B has been disturbed by its collision with the group, it may still retain a component of disk rotation. We find no gaseous connection between the tidal tails and NGC~7320C, but reaffirm the close connection between the shocked gas in the ridge (traced by X-ray, radio continuum and warm H$_2$ emission) and the southern tidal tail. Based on the integrated spectrum, we find a total \Hi\ mass in the group of 3.5$\pm$0.4$\times$10$^{10}$\Msol, higher than the VLA estimate and comparable to FAST.   
\end{abstract}

\keywords{Galaxy collisions (585); Galaxy interactions (600); Galaxy groups (597); Hickson compact group (729); HI line emission (690); Intergalactic medium (813); Intergalactic filaments (811); Shocks (2086)}

\section{introduction}
\begin{figure*}[!thbp]
    \centering
    \includegraphics[width=0.98\textwidth]{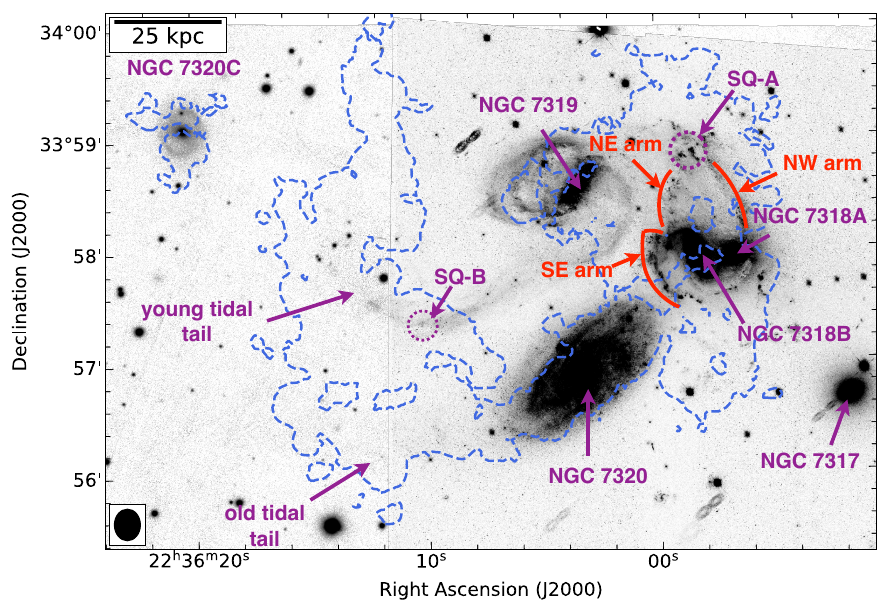}
 \caption{Outline map of the galaxies, stellar tidal features and \Hi\ distribution in Stephan's Quintet. The combined \textit{Hubble Space Telescope} (HST) and Digitized Sky Survey (DSS) optical image is overlaid with a MeerKAT \Hi\ contour drawn at $0.012 \rm \,Jy\,beam^{-1} km~s^{-1}$, corresponding to an \Hi\ column density of  $5.6\times10^{19}$ atoms\,cm$^{-2}$. Spiral arms and arm-like features in NGC~7318B are labeled in red. The 17\arcs$\times$14\arcs\ synthesized beam size is indicated in the bottom left corner. } 
      \label{fig::labeling}
\end{figure*}

Stephan's Quintet, also known as HCG~92 \citep{Hickson82} is perhaps the most famous compact group of galaxies, and is certainly among the most extensively studied galaxy systems on any scale. Originally identified as an optical grouping of five galaxies \citep{Stephan1877}, redshift measurements showed that one of the members (NGC~7320) is an unrelated foreground spiral galaxy. Of the remaining original members NGC~7317, NGC~7318A, and NGC~7319 form the core of the group (see Figure\,\ref{fig::labeling}) with recession velocities in the range $\sim$6600-6750\kmps \citep{Williams2002}. The last major galaxy, the highly inclined spiral NGC~7318B, appears to be falling through the group with a (blueshifted) relative velocity of $\sim$900\kmps.

The group shows many signatures of a complex interaction history, including tidal tails, arms, and optical filaments in and around the NGC~7318A/B pair and NGC~7319, see Figure\,\ref{fig::labeling}. Deep optical imaging shows an extensive filamentary halo of stellar emission surrounding and linking the galaxies \citep[e.g.,][]{Ducetal18,DuartePuertas2019}. However, the most spectacular feature of the group is a $\sim$50~kpc S-shaped ridge of shocked gas running along the east side of NGC~7318B. This was originally discovered from its radio emission \citep[][indicating the presence of relativistic electrons and a magnetic field]{Allen1972,vanderHulst1981} but also contains $\sim$0.6~keV thermal plasma visible in the X-ray \citep{Trinchieri2003,Trinchierietal05,OSullivan2009}, ionized gas traceable via H$\alpha$ emission \citep[e.g.,][]{Arnaudova2024}, molecular gas emitting via CO \citep[e.g.,][]{Emonts2025,Maedaetal2025,Xuetal25}, H$_2$ \citep{Appleton2023} and other lines \citep{Appleton2013}, as well as dust and star formation regions \citep[e.g.,][]{Xu2003}. The shock appears to have been driven by the collision of the intruder galaxy NGC~7318B with a pre-existing tidal filament of stripped gas \citep{Sulentic2001}, with the collision vector close to the line of sight.

Evidence for this earlier tidal structure has come primarily from \Hi\ line observations. Neutral hydrogen at the velocity of the group was first identified by \citet{Balkowski1973} and \citet{Shostak1974}, and the early interferometric \citep{Allen1980,Shostak1984} revealed gas components at roughly 5700, 6000 and 6600\kmps, including a curved tail extending several arcminutes east and north from the group, but little gas in the galaxies themselves. However, the detailed structure of the \Hi\ in the group was only fully established around the turn of the millennium, using Very Large Array (VLA) observations in C and D configurations (\cite{Williams2002}; see also \cite{VerdesMontenegro2001}). These revealed four main structures: the previously identified \Hi\ arc extending east and north of the group, overlapping the optical tidal tails but following a more tightly curved path; a low-velocity component in the southwest, apparently associated with the disk of the intruder galaxy NGC~7318B; and two north-western components, both located on a line of sight centered on the crossed northeast/northwest arms in the north of NGC~7318A/B, but with recession velocities $\sim$6000 and $\sim$6600\kmps. The highest column densities in the northwest components and in the arc were found to be associated with regions of star formation. One (SQ-A) is a collision triggered starburst \citep{Xuetal99,Xuetal25} the other (SQ-B) a tidal dwarf galaxy candidate \citep{Xuetal99,Lisenfeldetal04}. The radio continuum emission from the shock ridge was shown to fill the gap between the base of the \Hi\ arc and the northwest high-velocity component, suggesting the two had once formed an unbroken filament, with the \Hi\ in the ridge section since destroyed by the collision of NGC~7318B.

Later works have mainly focused on deep single-dish measurements in and around the group, seeking to measure the total mass of \Hi\ and identify any larger-scale structures. Additional diffuse \Hi\ not resolved by the VLA data is certainly present, with Green Bank Telescope (GBT) observations \citep{Borthakur2010,Borthakur2015} finding masses at least double that reported from the VLA, and Five-hundred-meter Aperture Spherical Telescope (FAST) mapping producing a total \Hi\ mass estimate of 3.45$\times$10$^{10}$\Msol\ \citep[for an assumed distance of 85~Mpc,][]{Xu2022,Cheng2023}. These observations also reveal \Hi\ emission at 6200-6300\kmps\ and 6800-7000\kmps which was not reported from the VLA data. Both the single dish and older interferometer datasets detected \Hi\ in other nearby galaxies not originally classified as part of the group, but whose velocities suggest they may be part of a halo of smaller satellites.

While more recent analyses of the archival VLA data have shown more extended \Hi\ structure in the group, with hints of gas overlapping NGC~7319 \citep{Jones2023}, it remains the case that the interferometric \Hi\ data for Stephan's Quintet is now more than 20 years old. We have observed the group using MeerKAT, with the goal of providing an updated view of its \Hi\ content and radio continuum emission. MeerKAT's 64 dishes provide a well-sampled $u-v$ plane with exceptional sensitivity to the L-band continuum, and \Hi\ line sensitivity and a maximum resolvable scale similar to that of the VLA ($\sim$10\arcm). In \citet[][hereafter paper~II], we will present an analysis of the multi-frequency radio continuum data for the group, including the MeerKAT L-band continuum maps. In this paper, we present the results of our \Hi\ analysis of the group.

The layout of this paper is as follows: in Section\,\ref{sec::data_reduction}, we present an overview of the observations and data reduction. The new \hi~images, overview of the detected structures, and integrated profiles and mass estimates are presented in Section~\ref{sec::results}. We end with a discussion and summary in Sections\,\ref{sec::discussion} and \ref{sec::summary}. 

Throughout the paper, we assume a distance to Stephan's Quintet (hereafter \SQ) of 94~Mpc, for easy comparison with recent studies of the group \citep{Appleton2023,Emonts2025,Aromaletal25,Xuetal25}, and use the optical velocity definition. At this distance 1\arcs\ corresponds to 456~pc. All output radio images are in the J2000 coordinate system and are corrected for primary beam attenuation.
\\


\begin{deluxetable}{lc}
\tablecaption{MeerKAT \hi~ observations detail \label{tab:obsparams}}
\tablewidth{0pt}
\startdata
\\
 Coordinates & RA: 22h35m57.5s \\
& Dec: +33d57'36.0" \\
Observation dates          & 19 December 2024 \\
                & 28 December 2024 \\
Flux/bandpass calibrator     & J0408-6545\\
Phase/gain calibrator        & J2236+2828 \\
Total integration time       & 6 h (2 $\times$ 3 h) \\
Correlator mode              & 32k, dual polarisation \\
Frequency range              & 0.9--1.7 GHz \\
Total bandwidth              & 856 MHz \\
Number of channels       & 32768 \\
Channel Width                  & 26.123 kHz\\
Spectral resolution          & 5.5 \kms \\
\enddata
\end{deluxetable}

\section{Observation and data reduction}
\label{sec::data_reduction}

The group was observed with MeerKAT using the L-band receiver and the 32k correlator mode, with an integration time of 8s.  For observational details, we refer to Table\,\ref{tab:obsparams}. The MeerKAT observations were centered at 1283\,MHz and had a channel width of 26.123~kHz, corresponding to a velocity resolution of 5.5 km/s. The large field of view of MeerKAT (HPBW $\sim$1\degree) enabled us to observe the group and its surroundings using a single pointing. However, due to its high declination, the source was observed in two separate observing runs ($3\times2$ hrs on source time). For each run, scans of the primary calibrator (J0408-6545) for flux/bandpass calibration were made at the start and end of the observations (10 minutes each). The secondary calibrator (J2236+2828) scans were conducted every 30 minutes for 2 minutes. Additionally, J0108+0134 was observed as a polarization calibrator. 

Each Meerkat observing run was processed using the Containerized Automated Radio Astronomy Calibration pipeline \citep[{\tt CARACal};][]{caracal2020}\footnote{\url{https://ascl.net/2006.014}}.  We proceed only with data covering the frequency range 1200-1520~MHz to limit the computing resources required for calibration and imaging. The data were calibrated on {\tt Hydra} (the Smithsonian Institution High Performance Computing Cluster). To produce the final \Hi\ cube, the data were calibrated in three steps: cross-calibration, continuum imaging, and line imaging.

\begin{deluxetable}{ll}
\tablecaption{MeerKAT H\textsc{i} Cube Properties\label{tab:H1_image_properties}}
\tablewidth{0pt}
\startdata\\
Frequency range        & 1370--1425 MHz \\
Velocity range & $-966.5$ to $11030.1\ \rm km\,s^{-1}$ \\
Beam (high resolution)   & $17\arcsec \times 5.5\arcsec$, $d\nu = 5.5\ \rm km\,s^{-1}$ \\
                         & rms $= 0.20\ \rm mJy\,beam^{-1}$ \\
Beam (medium resolution) & $17\arcsec \times 14\arcsec$, $d\nu = 16.5\ \rm km\,s^{-1}$ \\
                         & rms $= 0.18\ \rm mJy\,beam^{-1}$ \\
Beam (low resolution)    & $25\arcsec \times 25\arcsec$, $d\nu = 16.5\ \rm km\,s^{-1}$ \\
                         & rms $= 0.24\ \rm mJy\,beam^{-1}$ \\
\enddata
\tablenotetext{}{Notes: The high and medium resolution H\textsc{i} cubes are created with Briggs weighting with {\tt robust} $=-0.5$ while the low resolution one with {\tt robust} $=0.0$ without any taper. All reported rms values represent per-channel noise levels. $d\nu$ is the velocity resolution.}
\end{deluxetable}

\begin{figure*}[!thbp]
    \centering
    \includegraphics[width=1.00\textwidth]{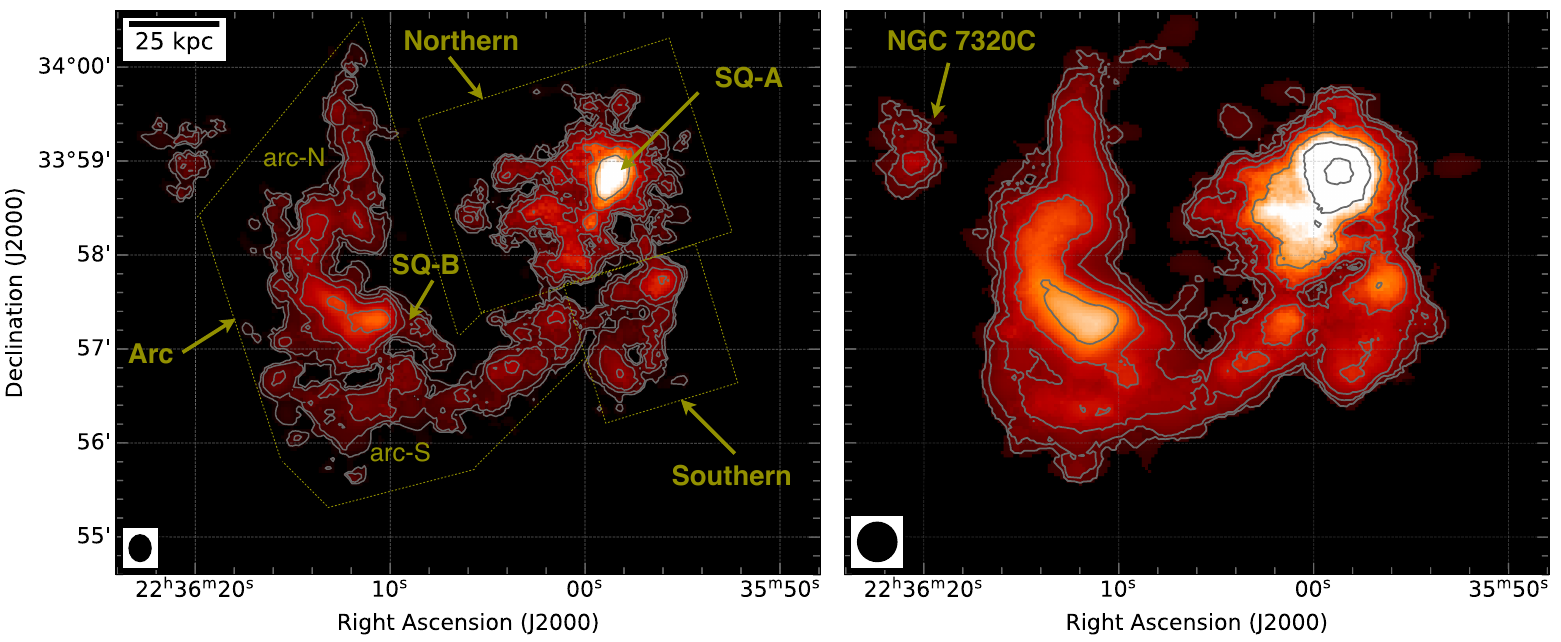}
 \caption{MeerKAT moment-zero maps at medium and low resolutions showing the total \Hi~column density distribution in \SQ. \textit{Left}: the beam size is $17\arcsec\times14\arcsec$ and the contours are drawn at $5.6, 11.1, 22.3, 44.5, 89.1, 178 \times 10^{19}\rm \,atoms\,cm^{-2}$. \textit{Right}: The beam size is $25\arcsec\times25\arcsec$ and the contours are drawn at $3.9, 7.8, 15.6, 31.1, 62.2, 124\times 10^{19}\rm \,atoms\,cm^{-2}$. The MeerKAT observations reveal previously undetected \hi emission in the group core, NGC~7319 and NGC~7320C.}
      \label{fig::mom0}
\end{figure*} 

\subsection{Cross Calibration}
The first step in {\tt CARACal} involved flagging shadowed antennas, autocorrelations, and known radio frequency interference (RFI) channels. Subsequently, {\tt AOflagger} \citep{Offringa2010} was employed to identify and remove additional low amplitude RFI-contaminated data. The frequency range covering Galactic emission was also flagged. 

The primary calibrator J0408-6545 was modeled using the MeerKAT Local Sky Models within {\tt CARACal}. Following this, cross-calibration was conducted to solve for time-dependent delays and gain calibration. We then applied the Bandpass correction, applied the primary calibrator delay and bandpass solutions to the secondary calibrator and solved for time-dependent amplitude and phase corrections. All cross-calibration solutions were applied to the target field, after which the target field was also flagged for RFI using the same  {\tt AOflagger}  strategy.

\subsection{Continuum Imaging}
After initial calibration, we created an initial image of the target field with {\tt WSClean} \citep{Offringa2014} within {\tt CARACal} using multi-frequency (MFS) synthesis. The continuum imaging was performed using the Briggs weighting scheme \citep{Briggs1995} with a robust parameter of $-0.5$ and uv taper of 10\arcsec. These parameters were chosen to achieve an optimal balance between angular resolution and sensitivity, with the uv taper specifically applied to recover extended continuum emission.  

For self-calibration, we employed {\tt CubiCal} \citep{Kenyon2018}.  After initial imaging, three rounds of phase-only self-calibration were performed, followed by a final round of amplitude-phase calibration.  During each self-calibration step, the {\tt SoFiA} \citep{Serra2015} source finder, integrated into {\tt CARACal}, was used to produce a sky mask from the output deconvolved images. The self-calibration solutions were interpolated in frequency and transferred to the cross-calibrated target measurement set.

\begin{figure*}[!thbp]
    \centering
    \includegraphics[width=0.8\textwidth]{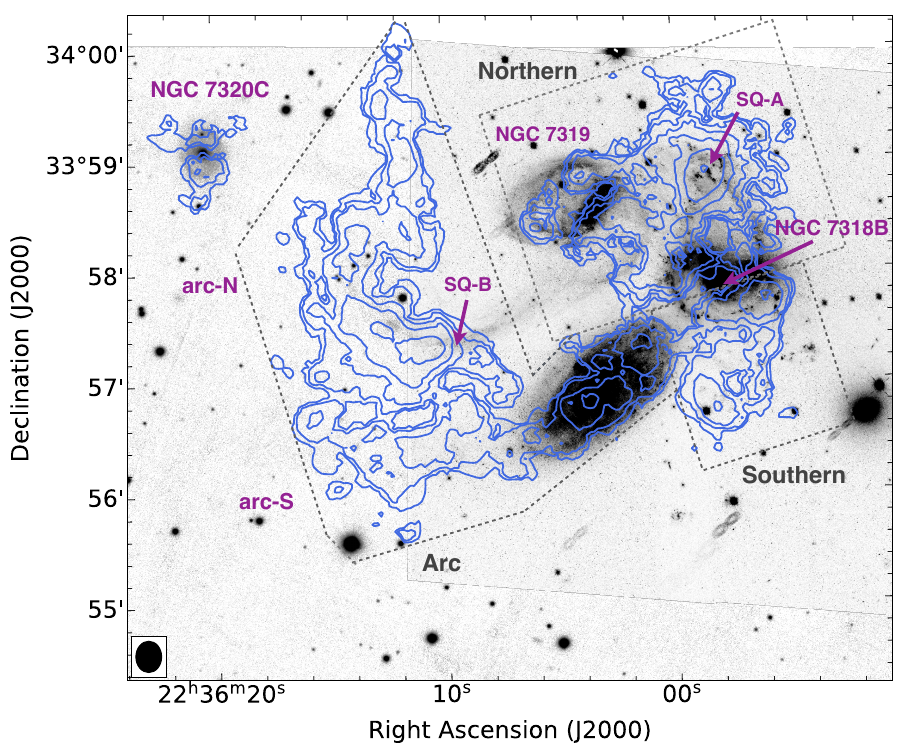}
 \caption{\textit{Top}: HST+DSS image overlaid with \Hi\ column density and radio continuum contours. The \Hi\ image beam size is $17\arcsec\times14\arcsec$ and the contours are drawn at $5.6, 11.1, 22.3, 44.5, 89.1, 178 \times 10^{19}\rm \,atoms\,cm^{-2}$ (left panel). }
      \label{fig::SQ_HI_labeling}
\end{figure*}

\subsection{Line imaging}
We subtracted the continuum model created in the previous step from the calibrated visibilities (both observing runs) using the task {\tt mstransform} in {\tt CARACal}.  Additionally, Doppler-tracking corrections and a fourth-order polynomial fitting to the visibilities are used to remove any residual continuum emission.  After this the \Hi\ data cube was produced with {\tt WSClean}, with a Briggs weighting with {\tt robust} value $r=-0.5$, multiscale cleaning, and without any uv-taper. This resulted in an angular resolution of $17\arcsec\times5.5\arcsec$. As noted above, the maximum resolvable angular scale with the MeerKAT is $\sim$10\arcm.  

Throughout this paper, we use two \Hi\ data cubes, referred to as the medium and low resolution cubes. Both cubes were smoothed in velocity, resulting in spectral-line maps with a velocity resolution of $16.5 \rm \,km\,s^{-1}$  (3-channel smoothing). The medium-resolution \hi~ cube has a synthesized beam of ($17\arcsec \times 14\arcsec$), obtained by smoothing the highest-resolution cube (i.e., $17\arcsec \times 5.5\arcsec$). This allowed us to make a direct comparison with the previously published VLA \Hi\ cube reported in \cite{Williams2002}. The single channel median rms noise in this cube is 180 $\mu$Jy\,beam$^{-1}$. The 3-sigma \Hi\ flux limit is about 0.009 Jy km/s, corresponding to an \Hi\ column density limit of $4\times10^{19}$ atoms\,cm$^{-2}$.  The low-resolution cube has a beam size of 25\arcsec, created with Briggs robust weighting (${\tt robust}=0.0$). The properties of the \Hi~ cubes are summarized in Table\,\ref{tab:H1_image_properties}. We also examined the MeerKAT data for evidence of OH line emission from the group, but none was found.

\section{Results: \Hi~ properties in Stephan's Quintet}
\label{sec::results}
Figure\,\ref{fig::mom0} shows our new MeerKAT total intensity \Hi~ emission in the direction of \SQ\ at the two spatial resolutions described above: $17\arcsec\times14\arcsec$ and $25\arcsec\times25\arcsec$.  The excellent sensitivity of MeerKAT allowed us to probe the \Hi\ emission with exceptional detail. The \Hi\ emission is mainly distributed in three distinct spatial regions: the Arc, Northern \citep[northwest high-velocity, northwest low-velocity in][]{Williams2002} and Southern \citep[southwest in][]{Williams2002}. Within the northern region, the strongest \Hi\ flux is seen from a relatively compact region roughly centered on the starburst region SQ-A. We recovered the \Hi\ associated with all previously identified features and robustly detected additional, previously unseen emission. The morphology and properties of the individual features are discussed in the following section.

\subsection{Overview of \Hi\ structure}
The most extended and diffuse \Hi~ emission is associated with the "Arc", a prominent L-shaped feature extending over 5.3\arcmin\ (about 140 kpc) east and southeast of the group (Figure\,\ref{fig::mom0}). In the MeerKAT maps, its morphology is consistent with that reported previously by \citet{Williams2002} and \citet{Jones2023} from the VLA observations.

\begin{figure*}[!thbp]
    \centering
        \includegraphics[width=0.49\textwidth]{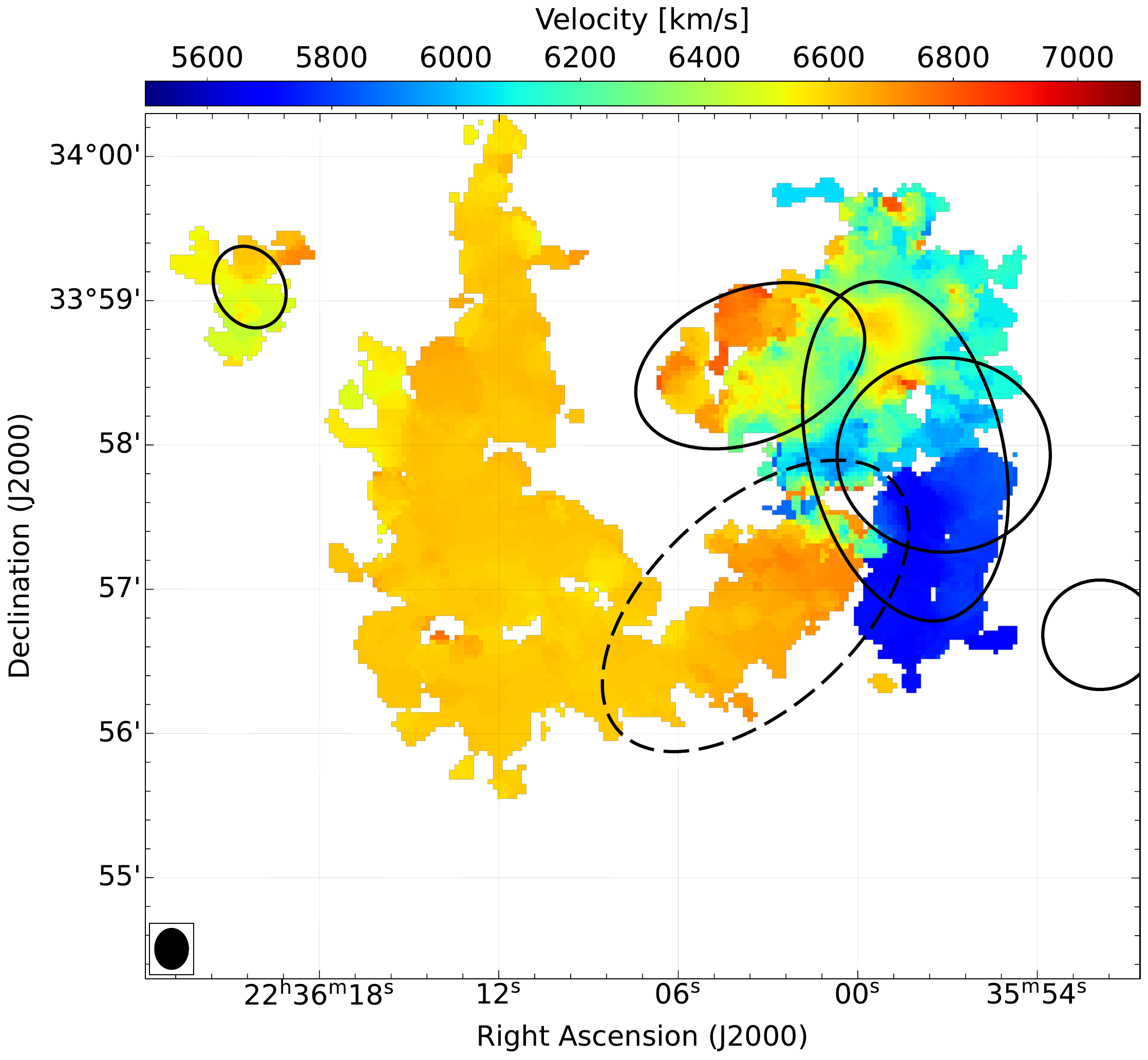}
    \includegraphics[width=0.49\textwidth]{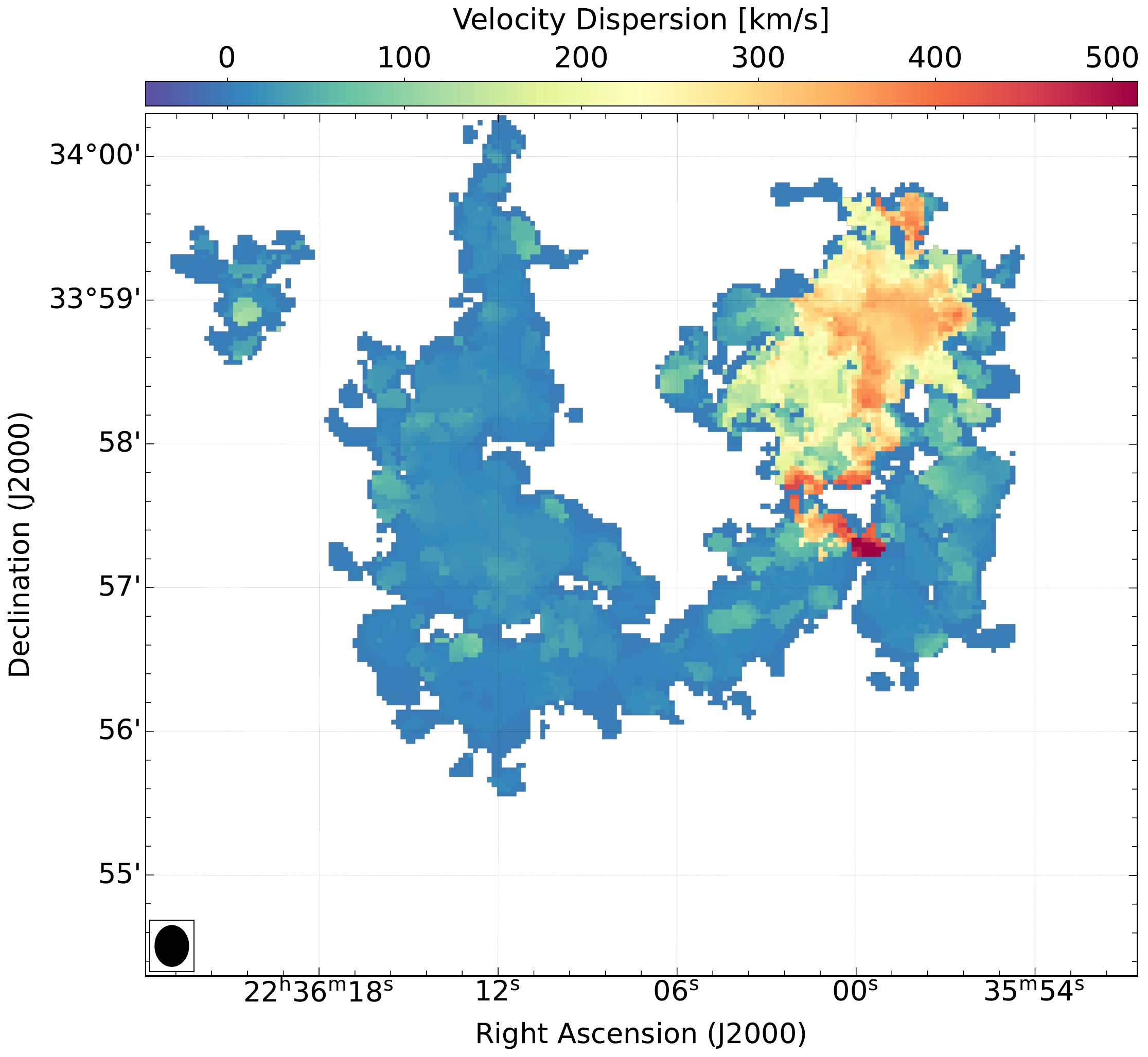}
 \caption{\Hi\ moment-one (Left) and moment-two (Right) maps of Stephan's Quintet obtained from {\tt SoFiA}. The solid black outlines mark the D$_{25}$ ellipses of the group member galaxies, using parameters from the LEDA catalogue, while the dashed ellipse marks the foreground galaxy}. These maps have a beam size of $17\arcsec\times 14\arcsec$. Beam sizes are indicated in the bottom-left corner of each image.
      \label{fig::SQ_maps}
\end{figure*} 

Figure~\ref{fig::SQ_HI_labeling} shows the \Hi\ distribution overlaid on an optical image (combined \textit{Hubble} and Digitized Sky Survey) of the group. The arc structure overlaps the two optical tidal tails, the northern young tail extending southeast from NGC~7319, which includes the star-formation region SQ-B, and the southern old tail extending southeast from behind NGC~7320. The arc has a distinct gap between the lower \Hi\ column-density gas in the southern tail (arc-S) and the higher column densities around SQ-B (arc-N), see Figures\,\ref{fig::mom0} and ~\ref{fig::SQ_HI_labeling}. Arc-N and arc-S appear to align with the optical tidal tails. \citet{Williams2002} posited such a separation, but the gap between the two components is considerably clearer in the MeerKAT data than it was in the VLA.

The arc has a mean radial velocity of about 6624\kmps\ and a very low velocity dispersion of about 10\kmps\ (see Figure\,\ref{fig::SQ_maps}). The radial velocity is comparable to that found with the VLA (6610\kmps). Unlike \citet{Williams2002}, we do not find any evidence for differences in kinematics or velocity dispersion between arc-N and arc-S. However, as in the VLA data, MeerKAT shows a velocity gradient at the base of arc-S, where it passes behind the foreground  spiral galaxy NGC~7320 \citep[v$_rec$=775\kmps,][]{Springobetal05}, with velocity increasing toward the northwest (Figure\,\ref{fig::SQ_maps}). Interestingly, in the L-band MeerKAT continuum image (PaperI), we find radio emission located along the northern optical tail but also between the two optical tails i.e., in the region separating arc-N and arc-S.

Within the arc, we identify two high column density clumps in the vicinity of SQ-B, though neither coincides with its position. The 25\arcs-resolution maps (Figure\,\ref{fig::mom0} right panel) show an additional spur of \Hi\ extending northwest along the base of the northern optical tail almost to NGC~7319, adding to the evidence that there are indeed two separate \Hi\ tails. To the east of the Arc lies the spiral galaxy NGC~7320C, which we detect in \Hi\ for the first time (see Figure\,\ref{fig::mom0}). We find no evidence of \Hi\ connecting arc-N and NGC~7320 but note that the gas velocities of the two are very similar (Figure\,\ref{fig::SQ_HI_labeling}).

A large fraction of the \Hi\ emission in the MeerKAT observation is detected in the bright northern region. Previous studies found this to consist of two components, both centered in the northern part of NGC~7318A/B around the starburst region SQ-A, but separated in velocity \citep{Williams2002, Shostak1984}. Our new observations reveal additional \Hi\ emission in this region, including the first clear detection of \Hi\ associated with NGC~7319. The northern components cover $\sim$2.5\arcmin\ east-west (corresponding to a physical size of $\sim70$~kpc) and extend significantly further south than was seen with the VLA, through much of the northern half of NGC~7318A/B and particularly along the eastern arms of NGC~7318B (see Figures\,\ref{fig::mom0} and \ref{fig::SQ_maps}). The peak \Hi\ emission is detected around the star-forming region SQ-A. By contrast, the nuclei of NGC~7318A and B show little to no \Hi\ emission, and the region immediately to their north, between the shock ridge and northwestern spiral arm is \Hi-free. As evident from the moment-one and two maps (Figure\,\ref{fig::SQ_maps}), the entire northern region spans a wide velocity range (6000-7000 \kms) and exhibits a high velocity dispersion (about 400~\kms).

We detect \Hi\ emission extending across much of the disk of NGC~7319 (Figure\,\ref{fig::SQ_HI_labeling}). The \Hi\ in the disk has an average radial velocity of $6686$\kmps, similar to but somewhat below the systemic velocity of $6740\pm50$\kmps\ derived from stellar absorption line measurements \citep{Aokietal96}. Nonetheless, this represents the first confirmed detection of \Hi\ in NGC~7319. The spatial distribution of the \Hi\ is broadly similar to the CO(2-1) molecular gas distribution found by \citep[][see Figure\,\ref{fig::H1_CO}]{Emonts2025}. CO emission peaks in the core of NGC~7319 (Seyfert-2 AGN) but while \Hi\ is present, no equivalent peak is found, indicating that the molecular-to-atomic mass ratio differs from core to disk, or that excitations conditions vary. 

The remaining \Hi\ emission in the direction of \SQ\ is found in a diffuse low velocity cloud located south of the cores of NGC~7318A/B, labeled as Southern region in  Figure\,\ref{fig::SQ_HI_labeling}. The morphology and extent of this feature are similar to that reported from the VLA data. It has a mean radial velocity of 5754\kmps, the lowest of any of the \Hi\ components, and a velocity dispersion of about 28~\kms (see Figure\,\ref{fig::SQ_maps}). It covers the southwestern part of the NGC~7318B SE spiral arm and extends considerably further south, overlapping a number of weaker stellar structures sometimes referred to as the "southern debris region" which may be part of the NGC~7318B disk. Notably, while the southern component abuts the northern component on the west side of NGC~7318A, in the east there is a distinct gap between the two.

\begin{figure*}[!thbp]
    \centering
\includegraphics[width=0.65\textwidth]{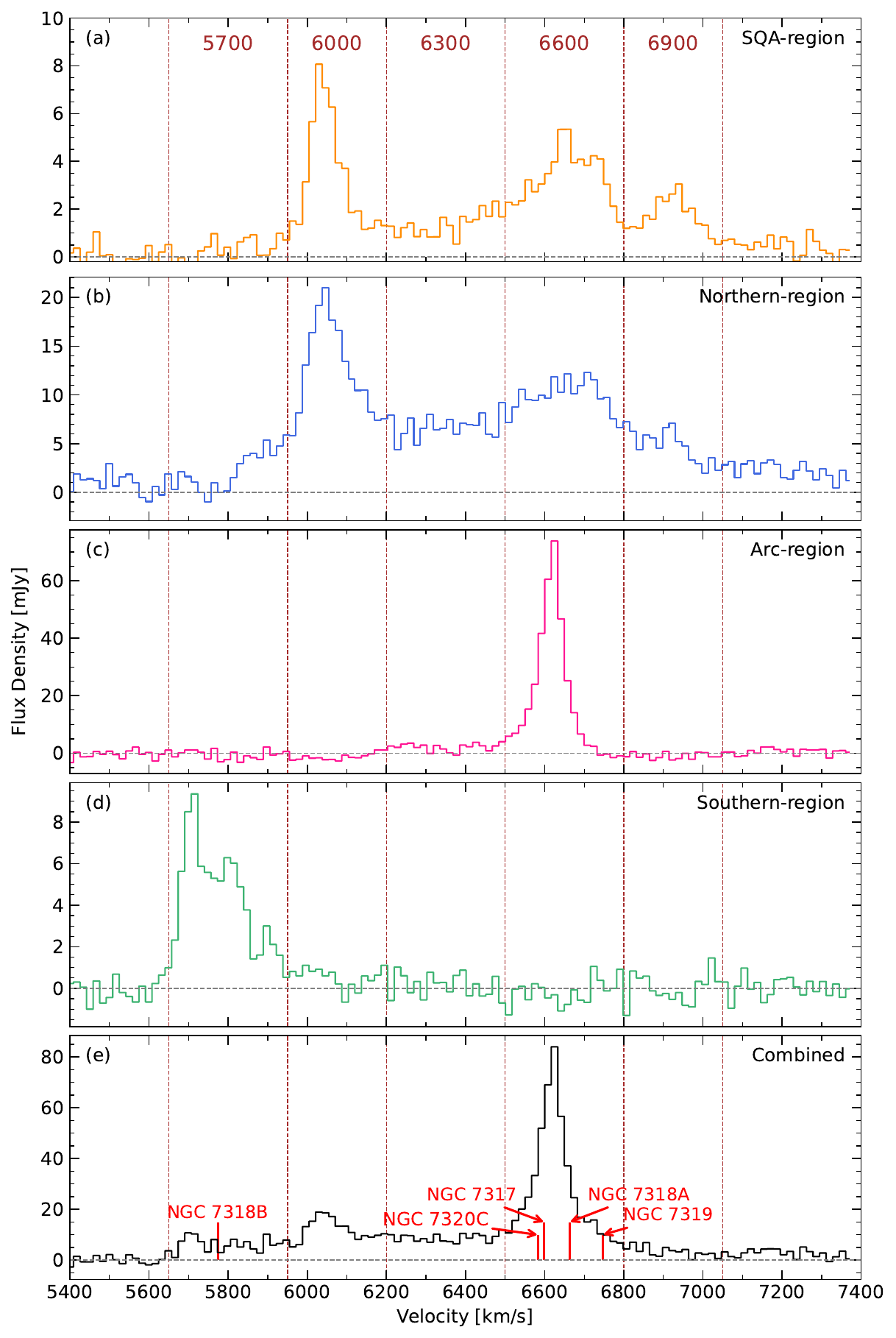}
 \caption{Spectra of the total \Hi\ emission and of individual subregions in \SQ. The brown vertical dashed lines mark the boundaries of the five kinematical components, namely 5700, 6000, 6300, 6600, and 6900. The dashed gray horizontal lines indicate the zero flux density level and red lines (with labels) mark the optical velocities of the major group member galaxies. Figure\,\ref{fig::HI_regions} shows the regions used to extract these. We note that the spectrum of the northern region includes the SQ-A region.}
      \label{fig::HI_Spectra}
\end{figure*}


\begin{deluxetable*}{l c c c c c c }
\tablecaption{\Hi~ Kinematical components of Stephan's Quintet
\label{table:H1_properties}}
\tablehead{
\colhead{Region} &
\colhead{Component} &
\colhead{Velocity range} &
\colhead{Integrated flux density\tablenotemark} &
\colhead{$M_{\rm HI}$ (MeerKAT)} & 
\colhead{$M_{\rm HI}$ (VLA)} &
\colhead{Figure} \\
\colhead{} &
\colhead{} &
\colhead{(km\,s$^{-1}$)} &
\colhead{(Jy km s$^{-1}$)} &
\colhead{($\times 10^{9}\,M_\odot$)} &
\colhead{($\times 10^{9}\,M_\odot$) }
}
\startdata 
 \multicolumn{7}{c}{Total} \\  
Integrated Spectrum & total&5400-7400 &$ 17.2 \pm 2.0$ & $ 35.0\pm4.0 $ & & Figure\,\ref{fig::HI_Spectra}\\
Spatially Resolved Maps &total&5400-7400 &$ 11.9 \pm 1.2$ & $ 24.0\pm2.0 $ &$13.4\pm1.6$ & Figure\,\ref{fig::SQ_HI_labeling}\\
   \hline
 \multicolumn{7}{c}{Main Structures} \\   
\multirow{3}{*}{Arc} & total &6200-6800  & $ 4.8 \pm 0.5$ & $ 9.8\pm1.0 $&$8.4\pm0.9$ & \multirow{3}{*}{Figure\,\ref{fig::SQ_HI_labeling}}\\
  &arc-N &6200-6800 & $ 3.2 \pm 0.3$& $6.5\pm0.6$ & \\
  & arc-S &6200-6800 & $ 1.6 \pm 0.2$& $3.3\pm0.4$ &\\
   \hline
Northern & total&5400-7400 & $ 6.0 \pm 0.6$ & $ 12.0 \pm1.0 $&$3.7\pm0.4$ &\multirow{2}{*}{Figure\,\ref{fig::SQ_HI_labeling}}\\
Southern &total &5400-7400 & $ 1.1 \pm 0.1$ & $ 2.4 \pm 0.3$ &$1.8\pm0.2$ \\
 \hline
 \multicolumn{7}{c}{Substructures} \\
\multirow{2}{*}{5700\, \& 6000} &total &5650-6200 & $3.2 \pm0.3$&$ 6.6 \pm0.7 $ & &\multirow{3}{*}{Figure\,\ref{fig::HI_a}}\\
\,\,components & 5700 & 5650-5950& $ 1.2 \pm0.2 $& $ 2.5 \pm0.3 $&\\
 & 6000 & 5950-6200& $ 2.0 \pm0.3 $& $ 4.1 \pm0.5 $&\\
  \hline
N7319 & &5400-7400 &$0.57\pm0.06$&$1.2\pm0.1$&& Figure\,\ref{fig::SQ_HI_labeling}\\ 
\hline
\multirow{4}{*}{SQ-A} &total &5400-7400 & $1.70 \pm 0.10$&$3.5 \pm0.4$ && Figure\,\ref{fig::SQ_HI_labeling}\\
  && 6800-7050&$0.20 \pm 0.02$ & $0.4 \pm0.04 $ && Figure\,\ref{fig::HI_b} \\
  && 6500-6800&$0.69 \pm 0.07$ & $1.4 \pm 0.1$ & & Figure\,\ref{fig::HI_b}\\
  && 5950-6200&$0.62 \pm 0.06$ & $1.3 \pm0.1 $ && Figure\,\ref{fig::HI_a} \\
\hline
SQ-B  &total& 5400-7400&$0.25 \pm 0.02$ & $ 0.50\pm0.04$ &\\
  \hline
\multirow{2}{*}{northern bridge} &total &5400-7400 & $0.36 \pm 0.04$&$0.73 \pm 0.08$\\
  && 6500-6800&$0.24 \pm 0.03$ & $0.49 \pm0.06 $& & Figure\,\ref{fig::HI_b} \\
    \hline  
\multirow{2}{*}{southern bridge} &total & 5400-7400& $0.38 \pm 0.04$&$0.78\pm 0.08$ & & \\
  && 6200-6500&$0.10 \pm 0.02$&$0.19\pm 0.02$ && Figure\,\ref{fig::HI_a}\\ 
\enddata
\tablenotetext{}{Notes: The total \Hi~ mass (i.e., Arc + Southern + Northern, excluding NGC~7320C) is obtained using two methods: (i) the flux density measured from the combined integrated spectrum, and (ii) the flux density measured from the \texttt{SoFiA}-generated pixel-wise maps at 25\arcsec\ resolution. The \Hi\ masses of the substructures are derived from the $17\arcsec\times14\arcsec$ resolution \Hi\ maps. VLA masses have been recalculated from the reported integrated fluxes using our adopted redshift and distance.}
\end{deluxetable*}

\subsection{Integrated \Hi\ Profiles}
\label{sec::int_profiles}

We extracted integrated \Hi\ line profiles for the four main regions described above (arc, southern region, northern region, and SQ-A). We used the medium-resolution MeerKAT \Hi\ cube ($17\arcsec\times14\arcsec$ beam size) to extract the spectra. The resulting integrated profiles are shown in Figure\,\ref{fig::HI_Spectra}. The regions from which the profiles were extracted are shown in Figure\,\ref{fig::HI_regions}. We note that SQ-A is a subset of the northern region, and the spectra of the latter include all flux density from the former. Past studies have shown considerable velocity structure around SQ-A \citep{Xuetal25}, and we thus consider it worthwhile to examine it separately as well as in combination with the larger-scale northern emission. 

The overall \Hi\ profile shape for the combined spectrum of \SQ\ is similar to those found with the VLA and FAST \citep[see Figure\,\ref{fig::FAST_MeerKAT};][]{Williams2002,Cheng2023}. The \Hi\ emission is detected over the velocity range 5500 to 7100\kmps. As with several previous studies \citep[e.g.,][]{DuartePuertas2019,Maedaetal2025}, we divide the spectrum into five velocity subcomponents, chosen based on the velocity structures in the integrated profiles: 5700 component (5650-5950 \kms), 6000 component (5950-6200 \kms), 6300 component (6200-6500 \kms),  6600 component (6500-6800 \kms), 6900 component (6800-7050 \kms). The strongest emission peak is seen at $\sim$6600\kmps, the second strongest at $\sim$6000\kmps, and we detect significant flux between these two as well as at more extreme velocities (Figure\,\ref{fig::HI_Spectra}). 

The three distinct emission peaks associated with the 5700, 6000, and 6600 velocity components (see Figure\,\ref{fig::HI_Spectra} panel e) have peak flux densities of 11 mJy, 19mJy, and 85 mJy respectively. These three components are reported in the VLA and FAST observations \citep{Williams2002, Jones2023, Cheng2023}. An additional weaker peak at higher velocities (6900 component) and lower-level intermediate-velocity emission (6300 component) were only detected in the FAST observations \citep{Cheng2023}. The distribution of the relative flux densities in our new MeerKAT \Hi\ profile is different from that of the VLA. For example, the peak flux density reported by \cite{Williams2002,Jones2023} for the 6600 component is lower by a factor of 1.5 than that found by MeerKAT. This discrepancy indicates additional \Hi\ emission recovered with MeerKAT, thanks to its excellent sensitivity and dense inner $u-v$ coverage. 

Compared to the single-dish FAST, GBT, and Arecibo data, MeerKAT recovers higher relative flux densities than GBT, values comparable to Arecibo, and lower values than FAST. Overall, the MeerKAT fluxes in the fainter 6300 and 6900 components are very similar to those found by FAST. However, the three main velocity components peaks (5700, 6000, 6600) detected with MeerKAT have peak flux densities lower than those from FAST (by a factor of 1.1-1.2). This difference can be partly attributed to resolution, as the FAST spectrum is obtained from a $2.9\arcmin$ resolution map. It is worth mentioning that NGC~7320C was not resolved by FAST, so flux from this galaxy is included in the FAST spectrum but not in Figure\,\ref{fig::HI_Spectra}. However, MeerKAT does resolve the galaxy, allowing us to treat it as a separate source.

The spectrum of the Arc structure shows a single narrow peak at $\sim$6600\kmps, as in all previous \Hi\ observations (Figure\,\ref{fig::HI_Spectra} panel c). Although it shows two distinct morphological features (arc-S and arc-N), the spectrum shows no evidence for multiple kinematic components. The southern region is similarly restricted to the low velocity 5700 component, again consistent with prior observations. However, unlike the Arc, in the southern region we find significant velocity substructure, with three velocity peaks (Figure\,\ref{fig::HI_Spectra}, panel d).

The dominant northern region shows the broadest profile with three distinct velocity components: 6000, 6600, and 6900 (Figure\,\ref{fig::HI_Spectra} panel b). This is the first detection of the 6900 component in interferometric \Hi\ data, and we measure an average radial velocity for this component of 6920\kmps. While \Hi\ emission from this 6900 component was previously found in single-dish data \citep{Cheng2023} it was associated with the group as a whole; MeerKAT now shows it to be primarily located in and around SQ-A (see Figure\,\ref{fig::HI_Spectra} panel a). This is consistent with molecular line observations \citep{Emonts2025}. The 6000 and 6600 components correspond to the NW-LV and NW-HV structures identified from the VLA data \citep{Williams2002} but we observe significant flux between these two peaks (the 6300 component) which was not seen by the VLA. We also see emission down to low velocities ($\sim$5800\kmps) in the northern region but notably not in SQ-A. 

From the combined MeerKAT spectrum, we estimate the total integrated line flux density $S_{\rm int}=\int S_{\nu} \, dv$ (in units of $\rm Jy \,km\,s^{-1}$) by summing $S_{\nu}$ covering the detected \Hi\ emission between 5500 and 7100 \kms. The uncertainties in the flux density reported in this work include the rms and the flux scale errors which we assume to be 10\%. From the spectrum, we obtain a total \Hi\ flux density of $17.2\pm2.0\,\rm Jy \,km\,s^{-1}$ for the group as a whole, which is 2.4 times higher than reported from the VLA \citep[7 $\rm Jy \,km\,s^{-1}$,][]{Williams2002} and 1.4 times higher than the GBT measurement \citep[12 $\rm Jy \,km\,s^{-1}$,][]{Jones2023}. Our measured total integrated flux density is slightly lower than the FAST value of $20.5\pm2\,\rm Jy \,km\,s^{-1}$  but consistent within the uncertainties (i.e., $1.2\sigma$ discrepancy). We note that the FAST total flux density includes an additional $0.2~\rm Jy \,km\,s^{-1}$ contribution from NGC~7320C. 

We also estimate the \Hi\ integrated flux density from \texttt{SoFiA} generated maps by applying 3-Dimensional masks on a channel by channel basis. The pixels outside the masks are set to zero. The advantage of this approach is that the channel-by-channel masks exclude pixels without \Hi\ emission and provide a higher signal-to-noise ratio. However, it may also exclude some pixels containing low-level emission, reducing the overall flux density measured for some regions. In Table\,\ref{table:H1_properties}, we summarize the MeerKAT \Hi\ global properties for \SQ\ and subregions. This method resulted in a total \Hi\ integrated flux density of $11.9\pm1.2$ Jy~km~s$^{-1}$, lower than that obtained from the \Hi\ spectrum but higher than reported by \cite{Williams2002} using the VLA. We emphasize that all integrated \Hi\ flux density measurements and \Hi\ mass estimates reported for sub-regions within \SQ\ are, unless otherwise stated, obtained from maps (medium and low resolution) generated with \textit{SoFiA}.

\subsection{Velocity substructures}
\begin{figure*}[!thbp]
    \centering
      \includegraphics[width=1.0\textwidth]{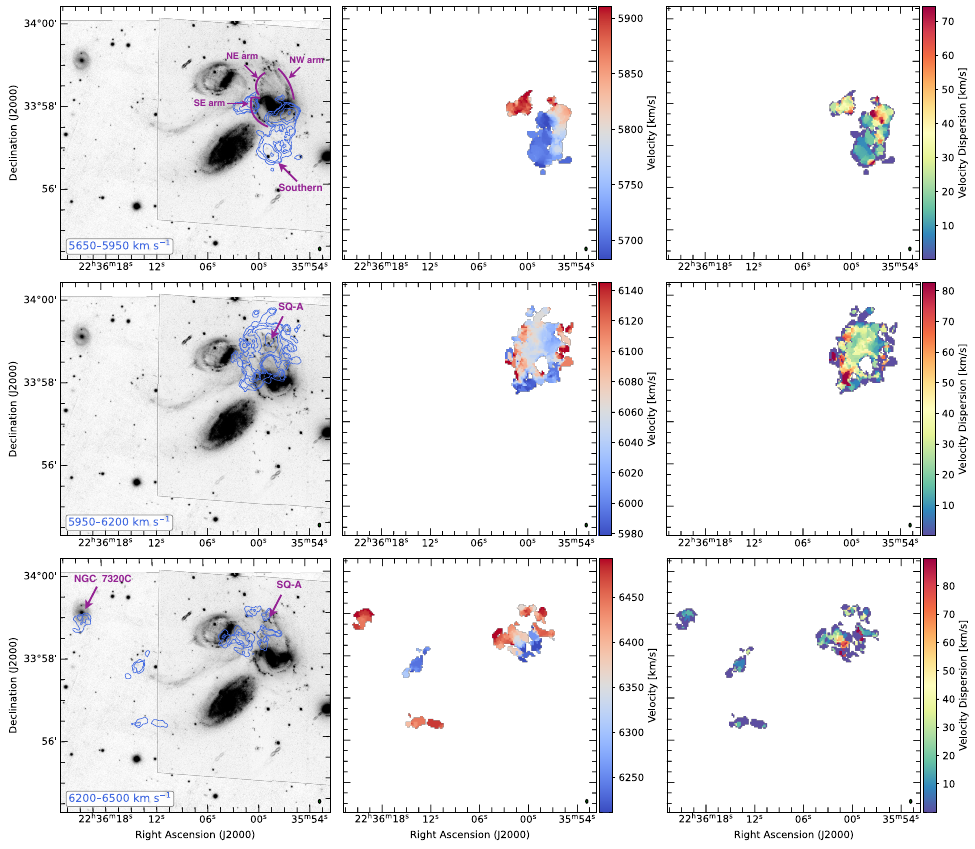}
 \caption{MeerKAT moment-zero (left), moment-one (middle) and moment-two (right) maps for the 5700, 6000 and 6300\kmps\ components. \Hi\ contour level are plotted at $5.6, 11.1, 22.3, 44.5, 89.1, 178 ...... \times 10^{19}\rm \,atoms\,cm^{-2}$. The radio beam size is $17\arcsec\times14\arcsec$.}
      \label{fig::HI_a}
\end{figure*}

\begin{figure*}[!thbp]
    \centering
      \includegraphics[width=1.0\textwidth]{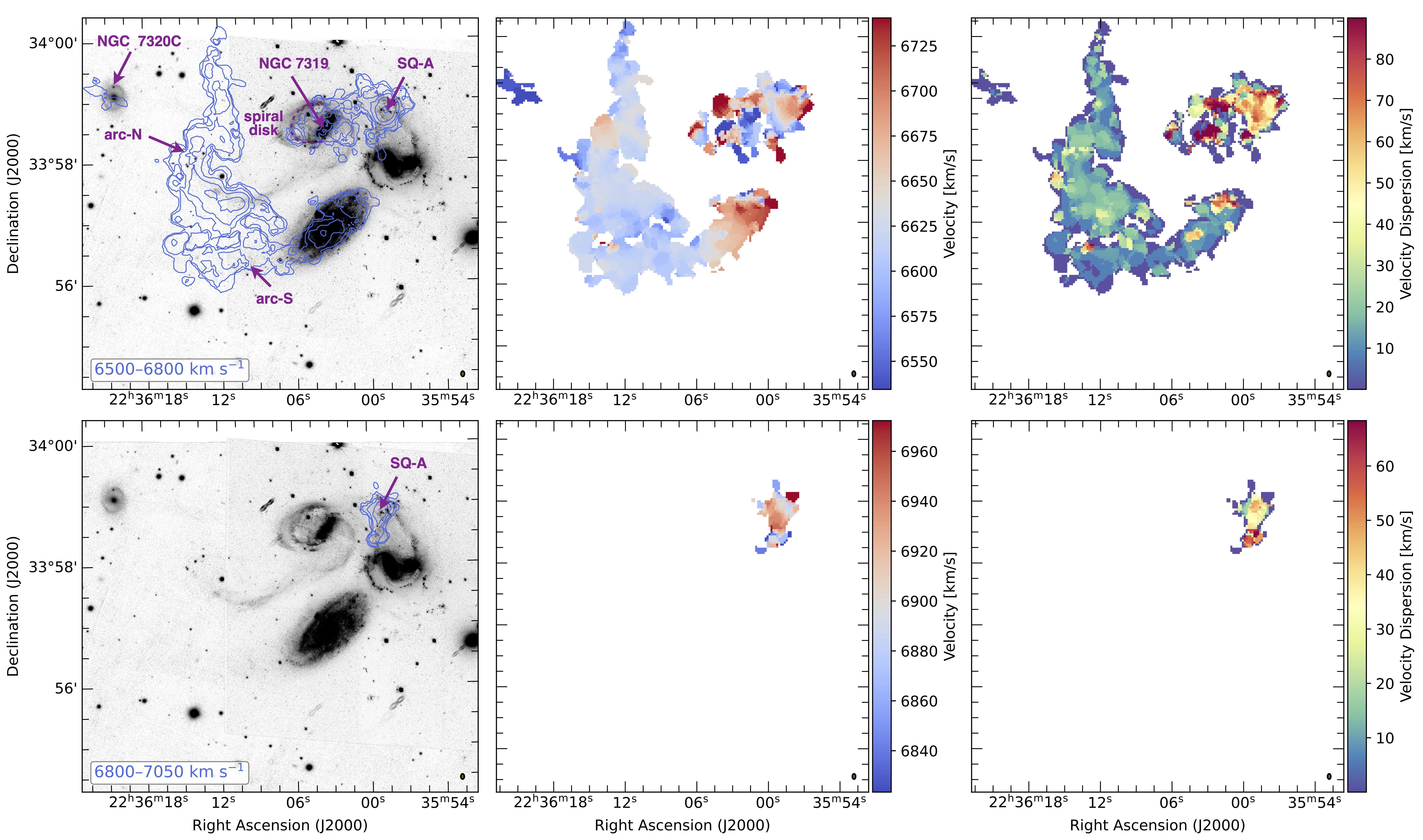}
 \caption{MeerKAT moment-zero (left), moment-one (middle) and moment-two (right) maps for the 6700 and 6900\kmps\ components. \Hi\ contour levels are plotted at $5.6, 11.1, 22.3, 44.5, 89.1, 178 ...... \times 10^{19}\rm \,atoms\,cm^{-2}$. The radio beam size is $17\arcsec\times14\arcsec$.}
      \label{fig::HI_b}
\end{figure*}

Figures~\ref{fig::HI_a} and \ref{fig::HI_b} show moment-zero, one,  and two maps for the five velocity ranges described above. These provide a more detailed view of the spatial distribution of the various velocity components and, in some cases, a clearer view of their kinematics. We also examined more finely-binned channel maps from the group, but given the wide velocity range of the \Hi\ emission and high spectral resolution of the MeerKAT data, we do not reproduce them here.


Figure~\ref{fig::HI_a} shows the low (5700, 6000) and intermediate (6300) velocity slices. The 5700 component shows two distinct clumps, the southern component discussed above and a smaller clump on the east side of NGC~7318B covering the shoulder of its southeastern spiral arm and extending toward NGC~7319 (top-panel). The latter overlaps emission in the 6000 component, which extends north from this position along the northeast arm (and shock ridge), peaks around SQ-A, and continues down the west side of NGC~7318A/B along the northwest arm (Figure~\ref{fig::HI_a} middle-panel). As noted above, this northern low-velocity component is considerably more extended than the VLA data suggested. The moment-one maps of the two low velocity components show evidence of east-west velocity gradients in both components, with higher velocities seen on the west side of the southern (5700) component, and on the east side of the northern (6000) component.

The intermediate velocity emission (6300 component) is resolved into a number of smaller clumps or agglomerations of \Hi\ (Figure~\ref{fig::HI_a} bottom-panel). These include gas in NGC~7320C, and small clumps in arc-S and arc-N coincident with the optical tidal tails but notably not associated with the star-formation regions of the northern (younger) tail. In the northern region, we see intermediate velocity gas near SQ-A, south of SQ-A in the northern part of the shock ridge, northeast of SQ-A, and extending west from the core of NGC~7319 across its disk toward the shock ridge. The moment one map suggests a gradient across this latter feature, which also overlaps lower-velocity gas at its eastern end (and higher-velocity gas in the west). This is the "bridge" structure reported across multiple wavebands, and whose velocity gradient was found from the ACA CO(2-1) data \citep{Emonts2025}.

Figure~\ref{fig::HI_b} shows the high-velocity (6600, 6900) components. As expected, much of the 6600 emission is found in the arc structure, with a relatively flat velocity distribution except at the base of arc-S, where there is a gradient with increasing velocities to the north (top-panel). We also see additional \Hi\ associated with NGC~7320C. In the northern component, the emission peaks around SQ-A, with extensions southeast along the shock ridge, northwest, and particularly to the northeast, where we see a second, northern bridge, linking SQ-A to the northwest side of NGC~7319. The \Hi\ associated with NGC~7319 is mainly found in the western disk (at the base of the southern bridge) and in clumps in the northeast and southeastern parts of the galaxy, a similar distribution to that found for CO(2-1) emission with ACA. The velocity structure of the northern component in complex, but in the 25\arcs$\times$25\arcs\ maps we see hints of higher velocities in the northeast quadrant of NGC~7319, and lower velocities in its southwest quadrant, again in agreement with CO(2-1). 

Finally, the highest velocity gas in the group (6900 component) is found in and around SQ-A, with an extension southeast along the shock ridge, and a stub extending east in the direction of the northern bridge linking across to NGC~7319 (Figure~\ref{fig::HI_b} bottom-panel). This is the most compact velocity component, but is still quite clearly aligned with the stellar structures in the shock ridge (the northeast arm).

\subsection{\Hi\ mass estimates}
We estimated the \Hi\ mass using the formula:
\begin{equation}
\frac{M_{\hi}}{M_{\odot}} =  \frac{2.36 \times 10^5 \,{D_L^2}}{(1+z)} \int S_{\nu} \, dv,
\label{eq: hi_mass}
\end{equation}
where $D_{\rm L}$ is the luminosity distance to the group in Mpc (i.e., 94~Mpc). For comparison with the \Hi\ mass estimate from the FAST observations, we follow \citet{Cheng2023} in using the integrated flux density from the MeerKAT combined spectrum (see Section\,\ref{sec::int_profiles}). This yields a total mass estimate of $3.5\pm0.4\times10^{10}\,M_{\odot}$ for \SQ, comparable to the FAST mass estimate of $4.2\pm0.5\times10^{10}\,M_{\odot}$. For a more direct comparison with the VLA mass estimate, we use the total (spatially filtered) integrated flux density measured from the \texttt{SoFiA} 25\arcs\ map (see Table~\ref{table:H1_properties}). This results in a lower \Hi\ mass estimate of $2.4\pm0.2\times10^{10}\,M_{\odot}$, but this is still a factor $\sim$1.8 greater than that reported \cite{Williams2002} using the VLA, reflecting the additional flux captured by MeerKAT.

Following \cite{Williams2002}, we estimate the mass in three main \Hi\ features in \SQ, namely the Arc, Southern, and Northern regions (see Figure\,\ref{fig::mom0}). The integrated flux densities and the corresponding \Hi\ masses of these regions are listed in Table\,\ref{table:H1_properties}. For the Southern region and Arc, we measured total \Hi\ masses of $2.4\pm0.3\times10^{9}\,M_{\odot}$ and $9.8\pm1.0\times10^{9}\,M_{\odot}$, respectively, slightly higher than found with the VLA. For the Northern region, our mass estimate is $12.0\pm1.0\times10^{10}\,M_{\odot}$, which is significantly higher than reported by \cite{Williams2002} (see Table\,\ref{table:H1_properties}). This reflects the considerably larger spatial extent of the Northern region in the MeerKAT data, the addition of the new 6900 component, and previously unseen flux at low and intermediate velocities (5700 and 6300 components).

To measure the \Hi\ mass of the smaller distinct regions, we used the medium-resolution maps, which better resolve and separate these individual structures. The atomic mass estimates for the northern and southern bridges are challenging because these features are detected over multiple velocity components (see Section~\ref{sec::discussion}). Therefore, we measure their masses both over the full velocity range and over the more restricted ranges where the bridges are clearly detected. Using this conservative approach, we constrain the mass of the northern bridge to the range $4.9\times10^{8}\,M_{\odot}<M_{\hi\,\rm Nbridge}<7.3\times10^{8}\,M_{\odot}$ and the southern bridge to $1.9\times10^{8}\,M_{\odot}<M_{\hi,\,\rm Sbridge}<7.8\times10^{8}\,M_{\odot}$. \cite{Emonts2025} reported a molecular gas mass for the southern bridge of $M_{\mathrm{H_2}} = (2.3 \pm 0.5)\times 10^8\,\alpha_{\rm CO}\,M_\odot$, and noted that the CO conversion factor $\alpha_{\rm CO}$ is unknown in \SQ, with plausible values ranging from $\sim$0.4 up to the canonical value for the Milky Way, $\alpha_{\rm CO}$=4.6. Nonetheless, this molecular gas mass is comparable to our estimated \Hi\ mass within uncertainties. This suggests that the molecular and atomic gas components contribute similar amounts to the total gas mass in this region. Such a molecular-to-atomic gas ratio is probably a little higher than that found in the outer disks of nearby spirals \citep{Leroy2008,Schruba2011} but as uncertainties are large and the southern bridge extends across the entire disk of NGC~7319, it is difficult to make a direct comparison.

For SQ-A, when considering the entire velocity range (for the used region see Figure\,\ref{fig::HI_regions}) we obtained a total mass of $3.5\pm0.4\times10^{9}M_{\odot}$. The \Hi\ mass of the individual velocity subcomponents range from $0.4-1.4\times10^{9}M_{\odot}$. The molecular mass of SQ-A is reported to be $M_{\rm H_2} = 0.10\,\alpha_{\rm CO}\ \times 10^{9}\ M_\odot$.

\subsection{Neighboring galaxies}
The MeerKAT observations also detect \hi~ emission from sixteen neighboring galaxies with radial velocities in the range 5400-7400\kms. Their column density, first-order and second-order moment maps are presented in Figures\,\ref{fig:appendix1} to \ref{fig:appendix4}. In Table\,\ref{tab:other_galaxies}, we summarize their properties and optical IDs. Among them, Anon~1, Anon~2, Anon~4, Anon~6, and Anon~8 were reported in previous observations \citep{Shostak1974, Williams2002, Xu2022, Cheng2023}. The remaining eleven sources in the \SQ~ field are either new detections, or new interferometric detections, having been previously identified in large single-dish surveys.

The MeerKAT derived radial velocities of Anon~1, Anon~2, Anon~4, Anon~6 and Anon~8 are consistent with previous studies \citep{Williams2002, Cheng2023}. For the newly detected source NGC~7320C, the radial velocity of 6520\kmps\ corresponds to a redshift of z=0.0218, in rough agreement with optical measurements from \citet{Sulentic2001} and \citet{DuartePuertas2019}, but conflicting with the $\sim$6000\kmps\ fiducial redshift in the NASA/IPAC Extragalactic Database (NED) and Lyon Extragalactic Data Archive (LEDA\footnote{http://leda.univ-lyon1.fr/}). The optical redshift of NGC~7320B (S1) implies a velocity of 6380\kms~ which is comparable to the \hi~ velocity of 6359\kms. Anon5 was reported by \citet{Shostak1974} as a single \hi\ source with a radial velocity of 6115\kms and a velocity width of 353\kms. In the same region, the MeerKAT data resolve the \hi~ emission into two separate sources (S3 and S9), both with optical counterparts. The majority of the galaxies show velocity distributions consistent with rotation, but S3 and S9 are asymmetric and may be \Hi\ tails. Sources S7 and S8 also appear somewhat disturbed, with multiple optical/IR sources within the \Hi\ contours, possibly indicating tidal interactions.

\subsection{Summary of key results}
As described above, the MeerKAT data reveals significant additional \Hi\ in \SQ\ compared to past interferometric observations, confirming the basic structure described by \citet{Williams2002} but identifying a number of important new features. New findings from these observations include:

\begin{enumerate}
\item The first detection of \Hi\ associated with member galaxies NGC~7319 and NGC~7302C,
\item Detection of significant additional emission in the north of NGC~7318B and between NGC~7318B and NGC~7319. In integrated maps, these galaxies are encompassed by a single $\sim$70~kpc wide \Hi\ structure, with total \Hi\ mass a factor $>$3 times greater than that found from the VLA.
\item Identification of the spatial location of the 6300 and 6900\kmps\ \Hi\ components previously found in FAST spectra. The 6900\kmps\ component is centered around the position of the SQ-A starburst, while the 6300\kmps\ component is found along lines of sight through the north of NGC~7318B and between NGC~7318B and NGC~7319.
\item Recovery of the majority ($\sim$80\%) of the \Hi\ emission found by single dish observations, suggesting that MeerKAT is likely capturing all the major \Hi\ structure in \SQ.
\end{enumerate}

In the following section we will consider these results in the context of past work on \SQ\ and examine their implications for the dynamics and internal structure of the group.

\section{Discussion}
\label{sec::discussion}

\subsection{Comparison with other wavelengths}
\begin{figure*}
\centering
\includegraphics[width=\columnwidth,bb=55 150 555 645,clip=]{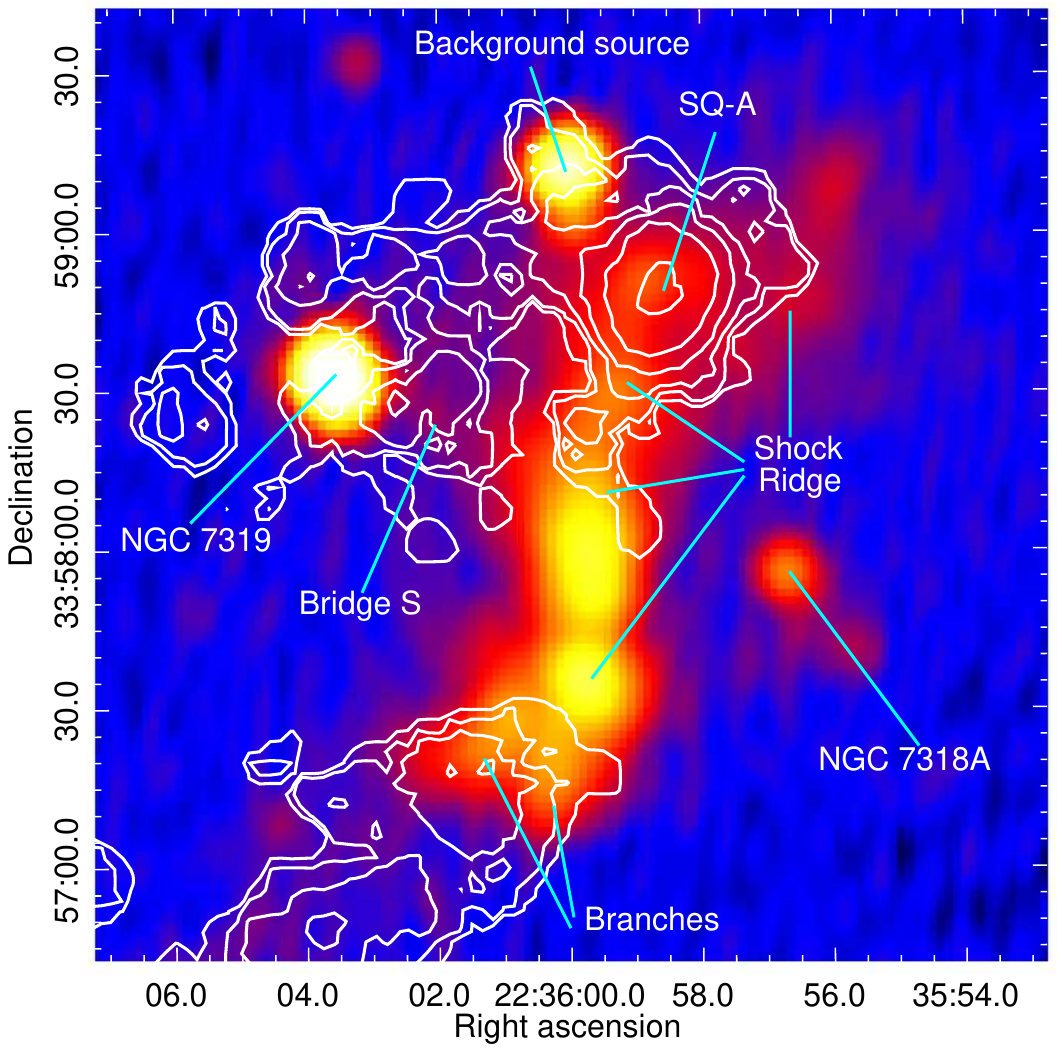}
\includegraphics[width=\columnwidth,bb=55 150 545 635,clip=]{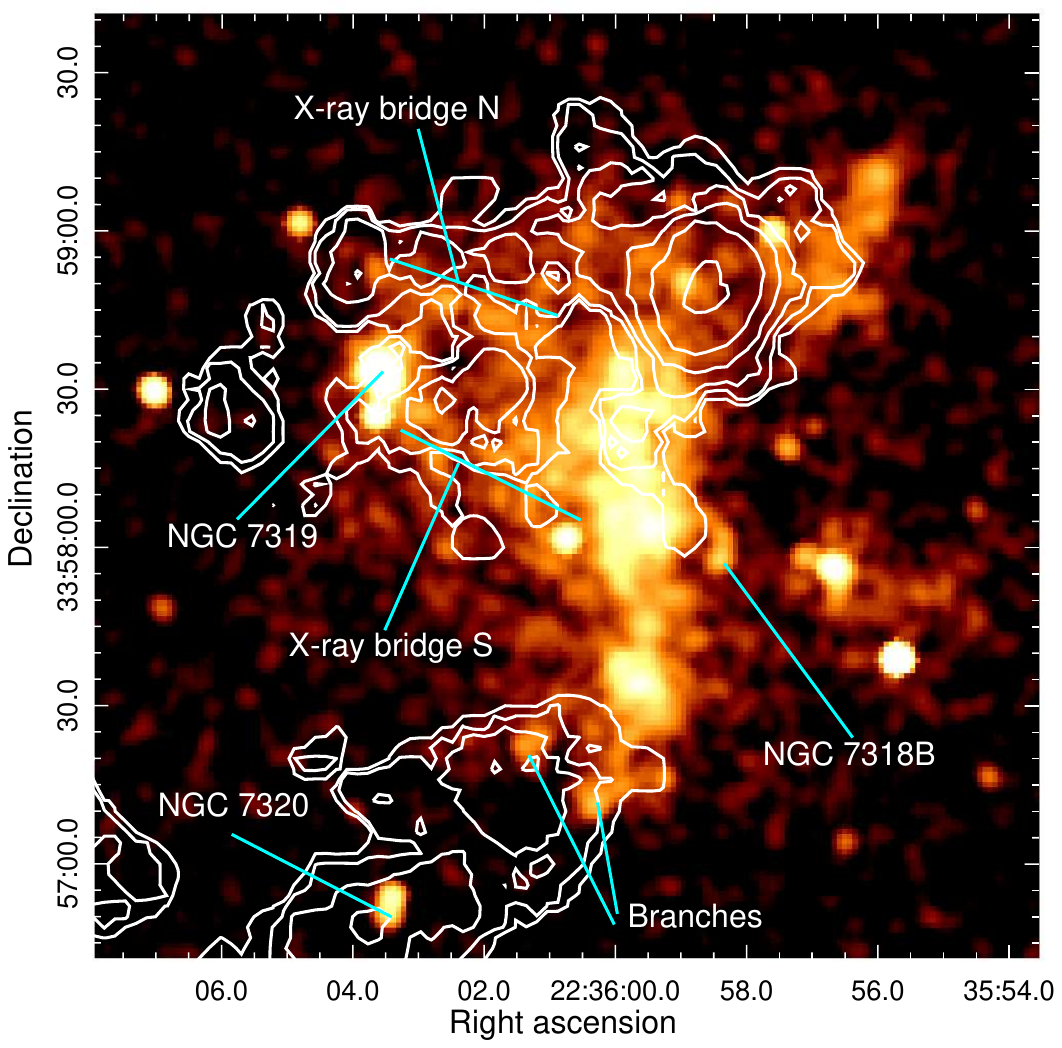}
\includegraphics[width=\columnwidth,bb=55 150 545 635,clip=]{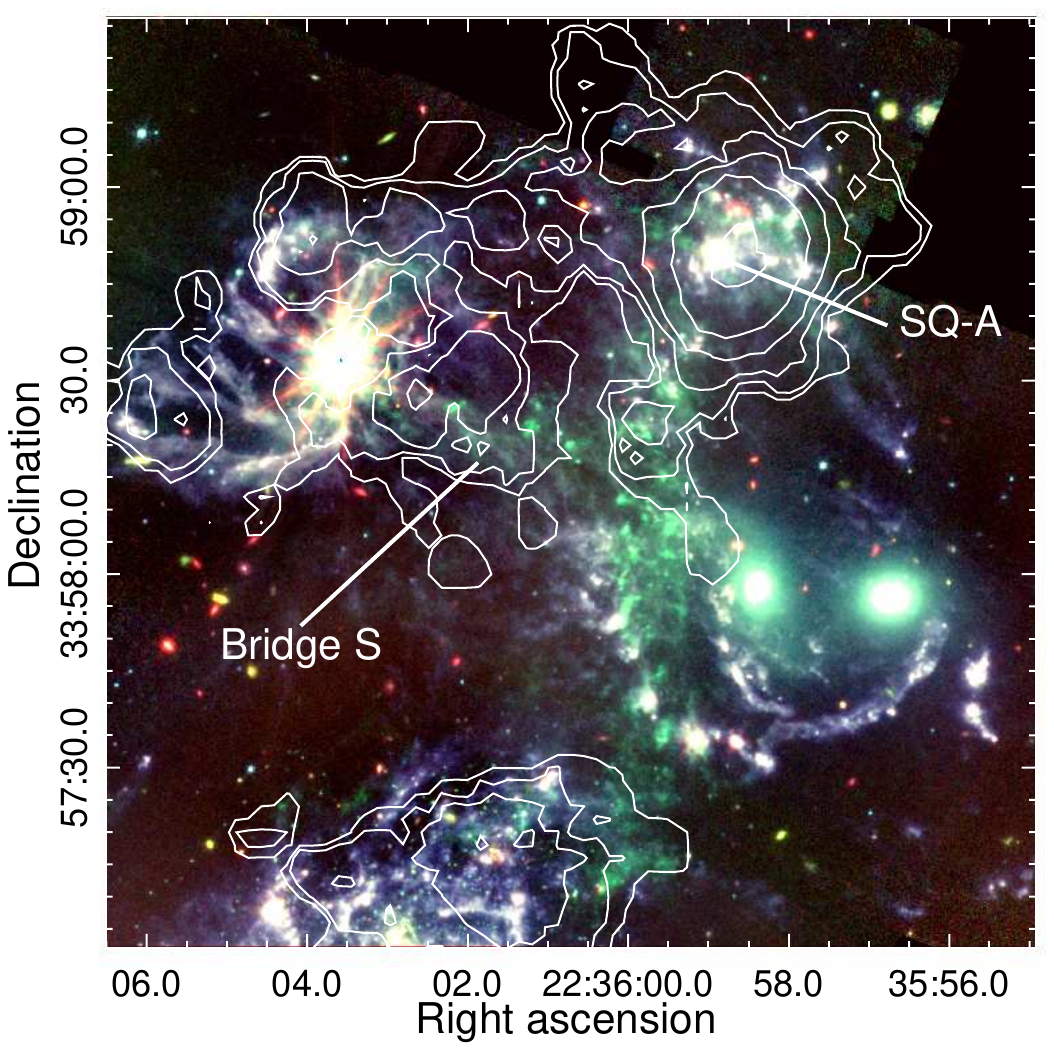}
\caption{\label{fig:Xray_JWST}\textit{Upper Left:} MeerKAT 8\arcs\ resolution L-band continuum image of the group core. Labels indicate main shock ridge, the active nuclei of two of the member galaxies, the southern bridge linking the shock to NGC~7319, and other features.\textit{Upper Right:} \chandra\ 0.3-2~keV image of the group core, smoothed with a 3\arcs\ Gaussian. Note the two filamentary X-ray bridges linking the shock ridge to NGC~7319. \textit{Lower:} \jwst\ MIRI false colour image of the group core with F1500W/F1000W/F770W images providing the red/green/blue channels. Note the green 10$\mu$m emission running vertically north-south along the shock ridge, and along the southern bridge to NGC~7319. All three images have MeerKAT \Hi\ 6500-6800\kmps\ contours overlaid.}
\end{figure*}

Given the additional \Hi\ identified by MeerKAT, it seems worthwhile to re-examine the relationship between this neutral atomic material and other structures in \SQ. Figure~\ref{fig:Xray_JWST} (upper-left panel) provides a comparison between the \Hi\ distribution in the 6600 component and the radio continuum emission in the group, dominated by the roughly S-shaped shock ridge. Previous studies established that this ridge connected the massive \Hi\ component around SQ-A to the base of the southern \Hi\ tail behind NGC~7320. This connection is thought to trace the line of a tidal gas filament that once included both the southern \Hi\ tail and the high-velocity ($>$6500\kmps) gas around SQ-A and NGC~7319. 

If we consider the summed MeerKAT \Hi\ map (Figure~\ref{fig::SQ_maps}), much of the gap in the \Hi\ emission along the shock ridge reported by, e.g., \citet{Williams2002} now appears to be filled in, with only a smaller \Hi-free region extending northwest from NGC~7320 toward the nucleus of NGC~7318B, crossing the southeast arm of the latter galaxy on its way. As shown in Figure\,\ref{fig::HI_a} top-panel, much of the new \Hi\ emission along the shock ridge is found at low velocities (5650-6200\kmps). This material seems to be associated with NGC~7318B and (in part) the southern bridge structure. At higher velocities, less \Hi\ is found in the ridge, and the southern half of the ridge appears to contain no \Hi\ at velocities $>$6200\kmps (Figures\,\ref{fig::HI_b} and \ref{fig:Xray_JWST}). Interestingly, the southern half of the shock ridge shows the brightest radio emission. A detailed comparison of the radio continuum emission with the \Hi\ is presented in paper~II.

Figure~\ref{fig:Xray_JWST} also shows comparisons of the 6600\kmps\ \Hi\ component with the shock region as traced in soft X-rays by \chandra\ and in the infrared by \jwst\ MIRI. The X-ray emission traces the same  S-shaped structure seen in the radio continuum emission: the main ridge runs north-south between roughly R.A.=22$^h$36$^m$00$^s$, Dec.=33$^\circ$58$^\prime$40$^{\prime\prime}$ and Dec.=33$^\circ$57$^\prime$30$^{\prime\prime}$, its northern end turns northwest across SQ-A, while the southern end of the ridge turns southeast and splits into two branches. These branches seem to wrap around the base of the \Hi\ tail (behind NGC~7320). The \chandra\ image also shows two filaments extending northeast from the ridge, one connecting to the nucleus of NGC~7319, the other toward its northern edge; we have labeled these as X-ray bridge south and north respectively. These two "bridge" structures have been identified previously from H$\alpha$ emission \citep[e.g.,][]{Sulentic2001,DuartePuertas2019} indicating the presence of warm $\sim$10$^4$~K ionized gas as well as the $\sim$10$^7$~K thermal plasma seen by \chandra. 

The \jwst\ MIRI \citep{Pontoppidanetal22,PontoppidanGordon22} bands all contain infrared continuum emission, but \citep[as discussed by][]{Appleton2023} the 10$\mu$m band (emission in green) also contains the relatively strong H$_2$ 0-0 S(3) emission line tracing warm molecular hydrogen. The 7.7$\mu$m band (emission in blue) has a weaker contribution from H$_2$ line emission and polycyclic aromatic hydrocarbon (PAH) emission, the latter being a good tracer of star formation. Blue and white regions, therefore, probably trace star formation and dust emission, while green regions are more likely to be dominated by H$_2$ emission. The 10$\mu$m is strong in the northern half of the ridge up to SQ-A, and in the southern bridge. Interestingly, in the southern half of the ridge, the 10$\mu$m emission follows the path of the radio continuum and X-ray emission, running north-south until it turns southeast and branches at the base of the \Hi\ tail. In contrast, emission in the 7.7 and 15$\mu$m bands traces the southeast spiral arm, crossing the ridge twice as the arm curves out eastward from the galaxy core, and then around to the southwest. This difference is also seen in the H$\alpha$, with the shocked, high velocity-dispersion gas following the north-south ridge, while the low velocity-dispersion emission from H\textsc{ii} regions follows the southeast spiral arm \citep{DuartePuertas2019,Arnaudova2024}. It therefore seems that the radio continuum, X-ray, H$_2$ and H$\alpha$ line emission are all tracing different aspects of the same shocked material.

The MeerKAT data shows the presence of \Hi\ along the line of sight to the northern part of the shock ridge at a range of velocities, including the high-velocity (6600, 6900) components which the pre-shock tidal filament may have connected to, and (in smaller quantities, closer to SQ-A) in the 6200-6500\kmps\ range that the shocked ionized component occupies (Figure\,\ref{fig::HI_a}). This suggests that at least some \Hi\ has survived the collisional shock or has been able to rapidly cool in this region, despite the harsh conditions. The southern bridge has been shown to contain both warm and cold molecular gas \citep{Appleton2023,Emonts2025}, the MeerKAT data shows it also contains \Hi, and we also see both X-ray and radio continuum emission. By contrast, the northern bridge is a much less coherent structure. The X-ray bridge runs along the inner edge of the \Hi\ bridge, neither component shows strong 10$\mu$m emission in the MIRI image, and while there is diffuse radio emission in the region, there is no clear separate radio structure associated with the bridge. The ACA did not detect CO(2-1) emission in the northern bridge, but new ALMA CO(2-1) observations do (Appleton, priv. comm.). We discuss the bridges in more detail in Section~\ref{sec:bridge}.

\subsection{The tidal tails and NGC~7320C}
The MeerKAT view of the \Hi\ tail structures agrees well with that found by the VLA, while adding some additional detail. It has been suggested that the optical and \Hi\ tails might have been drawn out by the passage of NGC~7320C through the group \citep{Molesetal98,Sulentic2001}, and deep optical imaging \citep[e.g.,][]{Ducetal18,DuartePuertas2019} reveals low surface brightness filaments connecting the optical tails (particularly arc-S) to NGC~7320C. Our MeerKAT observations reveal \Hi\ in the disk of that galaxy for the first time, and while it appears morphologically disturbed, it also shows a velocity gradient indicative of rotation, suggesting the gas is still bound. However, we do not see any direct connection to the \Hi\ tails. Instead, arc-N extends $\sim$2.5\arcm\ ($\sim$68~kpc) north of SQ-B, along a path traced by much fainter optical filaments.

The velocity distribution is flat over most of the arc, and the velocity dispersion is low (10-30\kmps) as expected for tidal structures (Figure\,\ref{fig::HI_b} top panel). We observe the same velocity gradient in the base of the arc-S (behind NGC~7320) reported by \citet{Williams2002} with velocities increasing to $\sim$6750\kmps. The tails are thought to have once been connected to the northern high-velocity (6600, 6900) \Hi\ components through what is now the shock ridge, and it is unclear whether this velocity gradient would have continued along this pre-existing filament. Without such a gradient, the tails would have connected to the more massive $\sim$6600\kmps\ northern component; with a constant gradient, the tails would instead connect to the $\sim$6900\kmps\ component. The presence of such a velocity gradient could have implications for the formation of the pre-collision filament.  

The highest column densities ($\rm 8.4\times10^{20}\,atoms/cm^2$) in the tails are found where arc-N overlaps the optical tail east of SQ-B. Previous CO observations have shown that $\sim$4$\times$10$^8$\Msol\ of molecular gas is located in SQ-B and the smaller star formation region at the end of the optical tail \citep[SQ-tip,][]{Lisenfeldetal02,Lisenfeldetal04}. This is comparable to the \Hi\ mass in the region, 5.0$\pm$0.4$\times$10$^8$\Msol. The CO mean velocity, 6625\kmps\ is in excellent agreement with the \Hi~  (see Figure\,\ref{fig::H1_CO}). We see no change in \Hi\ velocity structure in these regions; the mean velocity remains flat along the tail, and there is no obvious change in velocity dispersion. More diffuse molecular gas is likely present in much of the younger, northern tail \citep{Maedaetal2025} with a total mass of 1.85$\times$10$^9$\Msol\ (corrected to our adopted distance). This is around 30\% of the \Hi\ mass we find in the northern tail with MeerKAT.

\subsection{The low-velocity component and NGC~7318B}
Opinion has varied on the question of how much \Hi\ is still associated with the disk of the intruder galaxy, NGC~7318B. \citet{Molesetal97,Molesetal98} argued that the 5700 and 6000\kmps\ components found by the Westerbork Synthesis Radio Telescope were best interpreted as \Hi\ in the galaxy disk, with a central hole corresponding to the nuclear bulge of the galaxy. However, \citet{Williams2002} found a significant spatial gap ($\sim$15-45\arcs\ or $\sim$7-21~kpc) between the two \Hi\ components, as well as the 300\kmps\ difference in velocity. They argued that no plausible mechanism could produce such a separation, and concluded that only the 5700\kmps\ component was still associated with the galaxy disk.

MeerKAT shows that \Hi\ in the 5700 and 6000\kmps\ components does in fact cover much of the spatial extent of the disk of NGC~7318B, though there are still gaps in the galaxy core, the disk just north of the core, and in the southeast quadrant (see Figure\,\ref{fig::HI_a}, or Figure~\ref{fig::HI_regions} for a combined moment 0 map). The velocities of the two components overlap in the central regions where little \Hi\ was detected by the VLA and the overall velocity structure is consistent with disk rotation; we see the lowest velocities in the south, the highest in the north, and intermediate velocities to east and west of the galaxy core, with opposite velocity gradients in the north and south implying clockwise rotation. This would be consistent with the direction of curvature of the optical spiral arms. The relative velocity difference between north and south is $\sim$300\kmps. \citet{Yttergrenetal21} reported a stellar velocity gradient with a similar alignment, though their data only cover a northeast-southwest band centered on the galaxy core, and only show a relative velocity difference of $\sim$240\kmps. H$\alpha$ measurements show ionized gas at similar velocities to the \Hi\ in SQ-A, and in NGC~7318B's northeast and southeast arms \citep{DuartePuertas2019}. 

The \Hi\ spatial distribution is ellipsoidal, generally consistent with the optical and implying a disk with significant inclination to the line of sight. Adopting the optical inclination of 58.4\degree\ \citep{Makarovetal14} the true disk rotational velocity would be $\sim$175\kmps, not unreasonable for a relatively small spiral such as NGC~7318B. The total mass of \Hi\ in the low velocity (5700+6000) components is 6.9$\times$10$^9$\Msol, about 53\% of the predicted \Hi\ mass for the galaxy based on its luminosity \citep{Jones2023}. If we accept these components as a disk, the rotational velocity suggests a total gravitational mass of $\sim$3$\times$10$^{11}$\Msol\ within a radius of 41~kpc (1.5\arcm). The \Hi\ would thus make up $\sim$2\% of the total mass within this radius. 

However, there is a distinct offset between the \Hi\ velocities in the north and the optical recession velocity of the galaxy \citep[5774$\pm$24\kmps,][]{Hicksonetal92}. This could indicate that the disk rotation is asymmetrical or the gas is disturbed, or that the northern 6000\kmps\ component may not be associated with NGC~7318B. One possibility is that we are seeing the effects of a collision between the disk of NGC~7318B and the higher-velocity \Hi\ components along the line of sight through SQ-A. This collision might have driven gas out of the disk, decelerating it compared to the stars. It is notable that the gas in the 6000\kmps\ component follows the optical arms in the north of NGC~7318B, and thus must have interacted with the higher-velocity gas if NGC~7318B is moving along the line of sight.

\begin{figure*}[!thbp]
    \centering
\includegraphics[width=0.41\textwidth,bb=36 126 577 667]{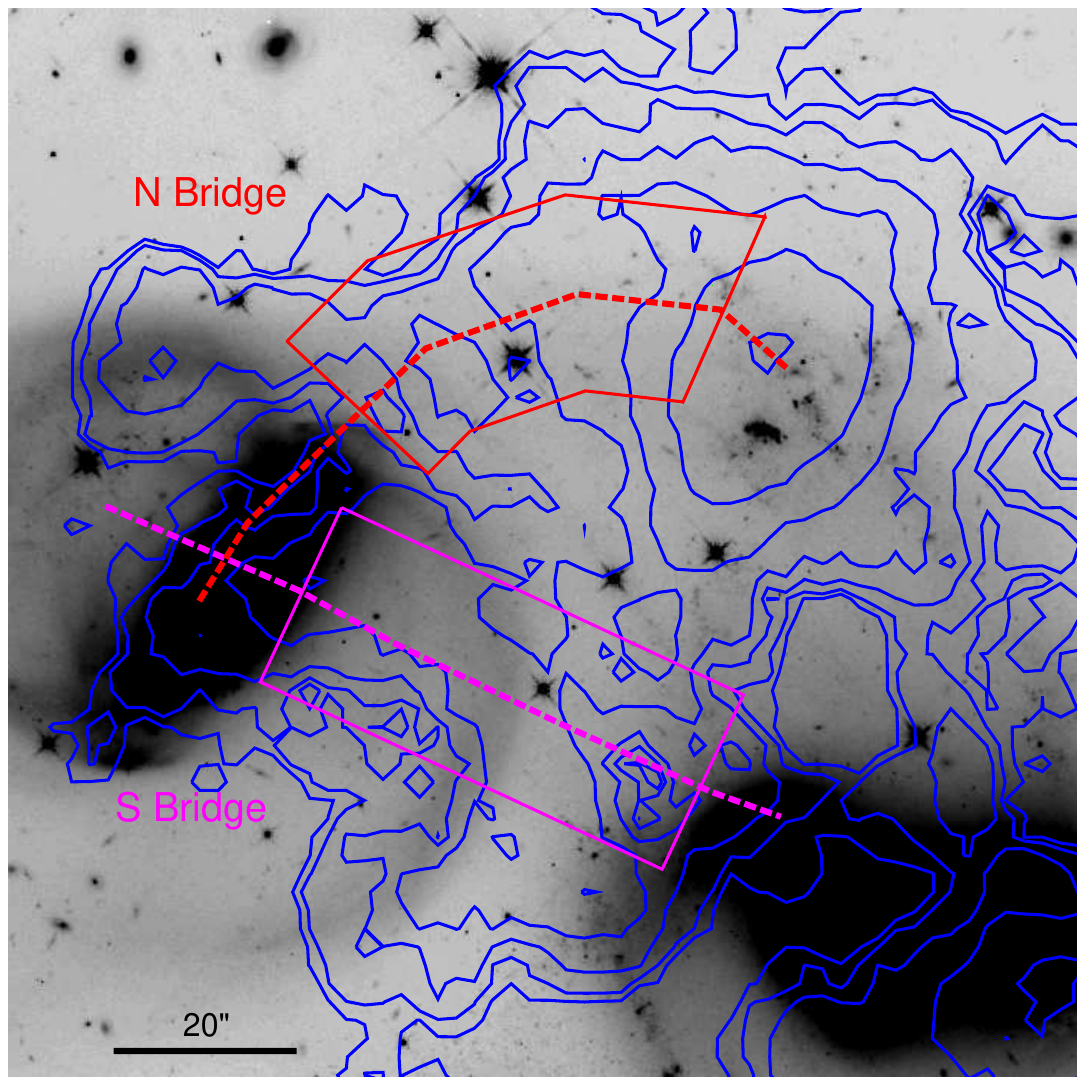}
\includegraphics[trim=65mm 0mm 0mm 0mm,clip,width=0.5\textwidth]{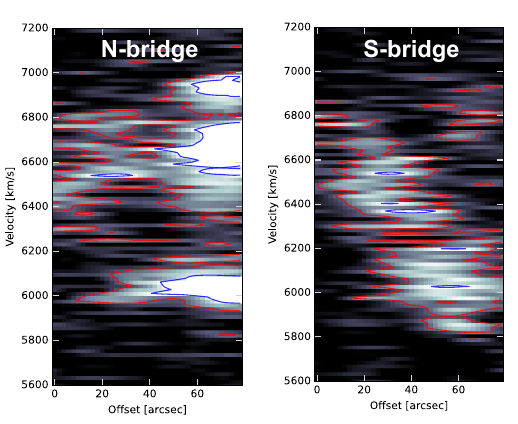}
 \caption{Segmented position-velocity diagrams of the northern (middle panel) and southern (right panel) bridges in HCG92, with red and blue contours showing \hi~ emission at 3 and 6 times the rms noise, respectively. In the left panel, dashed lines show the paths used to extract the PV diagrams, overlaid on the \hst\ F140W image, with total (5560-7050\kmps) \Hi\ contours overlaid in blue. Box regions show the 21\arcs\ width of the PV paths, and indicate the regions considered as bridges. In both center and right panels, the x-axis shows offset from east to west along these paths.}
      \label{fig::pv-diagram}
\end{figure*}

\subsection{The high-velocity components: NGC~7319 and SQ-A}

As noted previously, the VLA observations found the northern $\sim$6600\kmps\ \Hi\ component to be centered around the SQ-A starburst region, with no emission observed in NGC~7319 or between the galaxies \citep{Williams2002}. The MeerKAT data show significant additional structure (Figures\,\ref{fig::HI_a} and \ref{fig::HI_b}). Firstly, the \Hi\ associated with SQ-A is now revealed to have a previously unknown, relatively compact $\sim$6900\kmps\ component. Secondly, the known $\sim$6600\kmps\ component is much more extended than the VLA data suggested, with significant emission in the disk of NGC~7319 and a bridge linking the north-western edge of that galaxy to SQ-A (Fig.~\ref{fig::HI_b}). 

In general, the distribution of the high-velocity \Hi\ components in the northern part of the group are suggestive of tidal structures. While \Hi\ is seen in NGC~7319 for the first time in modern interferometric observations, its distribution is very lopsided with far more gas on the west side of the galaxy. The newly-identified northern \Hi\ bridge linking it to the SQ-A region as well as the previously identified southern bridge linked to NGC~7318B may also suggest that cool gas has been transported westward from the galaxy by tidal interactions. Tidal stripping is common in compact groups \citep[e.g.,][]{Verdes-Montenegro2001,Konstantopoulos2010,Borthakur2010,Jones2023} and these new structures may well have formed in the interactions which produced the tidal tails. Ram-pressure stripping generally has less impact in low-mass groups \citep{Roberts2021} than in galaxy clusters and is unlikely to be effective on short timescales in the relatively limited IGrM of SQ, except possibly for gas in NGC~7318B, which is moving supersonically compared to that IGrM \citep{OSullivan2009}. The peak electron density in the shock ridge is $\sim$2.5$\times$10$^{-2}$\pcmcu\ \citep{OSullivan2009}, but in the surrounding halo densities will be at least two orders of magnitude lower. The hot IGrM is also rather irregular and only detected out to $\sim$100~kpc \citep{Trinchierietal05,OSullivan2009}.

In NGC~7319, the velocity of the \Hi\ overlaps the optical recession velocity of the galaxy \citep[6747$\pm$4\kmps,][]{Nishiuraetal00} but the mean \Hi\ velocity is $\sim$100\kmps\ lower. In the 25\arcs\ resolution maps we see some indications of a velocity gradient, with velocities of $\sim$6750\kmps\ on the northeast side of the disk and $\sim$6550\kmps\ on the southwest. This could indicate rotation, but given the evidence of tidal disturbance of the galaxy, and the bridge of \Hi\ which links the southwest quadrant of NGC~7319 to NGC~7318B, it seems more likely that the lower velocities arise from unbound or disturbed gas. \citet{Yttergrenetal21} found an even stronger gradient in ionized gas emission along a similar axis, but noted that the velocity gradient of the stellar component is offset by almost 90\degree\ and that the centers of the stellar and gas velocity fields are offset.

\citet{Emonts2025} report CO emission from the nucleus of NGC~7319 and extending along the southeast spiral arm or loop, another clump in the northeast part of this loop, and extending into the bridge structure (discussed below). We see \Hi\ clumps on the eastern side of the galaxy overlapping the regions detected in CO, and on the west side extending from the galactic bar down to the bridge (Figure\,\ref{fig::H1_CO}). The CO emission shows the same velocity gradient we observe in the \Hi, providing evidence that the two gas phases are co-spatial. However, we do not see strong \Hi\ emission to match the CO emission from the nuclear region. The bright Seyfert nucleus of NGC~7319 may have played a role in driving out or ionizing low-density atomic gas from the nuclear region \citep{Aokietal96,PereiraSantaellaetal22}. The total molecular mass in the galaxy depends on the assumed CO conversion factor $\alpha_{\rm CO}$. For the typical value found in the Milky Way \citep[$\alpha_{\rm CO}$=4.6,][]{SolomonVandenbout05} the molecular gas mass would be $\sim$2$\times$10$^9$\Msol, roughly double the mass of \Hi\ observed in the galaxy. This could indicate that past interactions have been more effective in removing low-density material, perhaps because of differences in distribution; a more extended low-density gas disk might have proven more susceptible to tidal forces. Alternatively, interactions may have caused compression of the \Hi, leading to the formation of more molecular gas. However, the true value of $\alpha_{\rm CO}$ in NGC~7319 is unknown, and the molecular gas mass could be significantly lower.

\subsection{The bridge structures}
\label{sec:bridge}

As noted above, two \Hi\ bridge structures link NGC~7319 to NGC~7318B and the SQ-A region. Components of both are visible at multiple frequencies, indicating highly multi-phase gas, as in the main shock ridge. While all components of the southern bridge follow a consistent path, in the northern bridge there are differences. The stellar component follows a curved path, with multiple arcs linking the northwest side of NGC~7319 to the northern part of NGC~7318B and SQ-A. In contrast, the northern X-ray bridge and H$\alpha$ emission seem to follow separate but straighter lines, with the X-ray bridge (Figure~\ref{fig:Xray_JWST}) running roughly southwest-northeast, and the relatively weak H$\alpha$ running east-west \citep{Sulentic2001,DuartePuertas2019}. The \Hi\ bridge is primarily seen in the 6600 component, linking the region around SQ-A to the northwest edge of NGC~7319 and the \Hi\ in the disk of that galaxy. While the emission from the different gas and stellar components overlaps, there are distinct offsets in position, and without velocity measurements we cannot be sure whether the ionized gas and X-ray plasma are cospatial with the \Hi. The position-velocity (PV) diagram for the northern bridge (Fig.~\ref{fig::pv-diagram} middle-panel) shows no strong evidence of a velocity gradient. In the center of the bridge the \Hi\ covers a velocity range $\sim$6500-6800\kmps, somewhat narrower than the velocity ranges in the galaxies, and there seems to be no connection to the 6900\kmps\ component in SQ-A. The 6000\kmps\ in northern NGC~7318A/B overlaps the northern bridge but does not extend into NGC~7319. We note that using an east-west route for the northern bridge, between SQ-A and the northeastern-most \Hi\ clump in NGC~7319, would not change our conclusions; it shows the same mid-bridge velocity range and the same $>$3$\sigma$ significant connection at velocities $\sim$6780\kmps, but the northeast clump lacks the emission at $\sim$6680 and $\sim$6540\kmps\ we see in the central regions of NGC~7319.

The southern bridge has been detected in H$\alpha$ \citep[e.g.,][]{Sulentic2001}, H$\beta$ \citep{Guillardetal22}, warm H$_2$ \citep{Cluveretal10,Appleton2017}, [CII] \citep{Appleton2013} and CO \citep{Guillard2012} as well as the X-ray and radio continuum emission noted above. \citet{Emonts2025} revealed a strong velocity gradient in the ACA CO(2-1) data for the bridge, with velocities falling by $\sim$700\kmps\ over a distance of $\sim$25~kpc from NGC~7319 to the shock ridge. Ionized gas clumps are observed in the bridge at $\sim$6400\kmps\ \citep{Konstantopoulos2010,DuartePuertas2019}, consistent with the CO velocities. Figure.~\ref{fig::pv-diagram} right-panel shows the \Hi\ position-velocity diagram along this structure, though covering a broader (21\arcs) band than the CO owing to the larger \Hi\ beam size. We see the same overall gradient, with velocities falling from $\sim$6600\kmps\ in the core of NGC~7319 to $\sim$5900\kmps\ at the shoulder of the southeast spiral arm of NGC~7318B. In the CO, separate components are seen in the NGC~7319 nucleus and eastern bridge, merging to form a single structure in the western bridge. We do not observe this in the \Hi, and while the broader \Hi\ beam size might make separation of nucleus and bridge difficult, we see no hints of a broader \Hi\ velocity distribution, as might be expected. We do observe some weak (3$\sigma$ significant) emission clumps apparently unrelated to the bridge in the $\sim$6700-6900\kmps\ range on the western side. These correspond to the  small quantities of \Hi\ at the group velocity which extend down the shock ridge from SQ-A into the shoulder of the southeast spiral arm, as seen in Fig.~\ref{fig::HI_b}.

The origin of the two bridges is unclear, and indeed their properties may suggest different formation histories. In the southern bridge, the close correlation of components with very different temperatures suggests that the bridge was originally a cool gas structure which has been shock heated, as is the case for the main shock ridge. However, the fairly strong velocity gradient of the bridge is difficult to explain. While the Seyfert nucleus of NGC~7319 has been shown to be driving outflows \citep[e.g.,][]{Aokietal96} and radio jets \citep{Xanthopoulos2004} aligned roughly southwest-northeast, their scale ($\sim$2-5~kpc) is much smaller than the bridge and the jet alignment is offset from the bridge by $\sim$40\degree. It is also unclear how an outflow could produce the observed velocity gradient, or why it should connect to the velocity of NGC~7318B. At first glance a tidal origin for both bridges seems more plausible, with gas being drawn out of NGC~7319 by the interactions that formed the high-velocity \Hi\ components around SQ-A, the older (southern) tidal tail, and the filament which is presumed to have linked them. The relatively flat velocity across the northern bridge would be consistent with an interaction primarily occurring in the plane of the sky. However, the velocity gradient in the southern bridge would then have to be the product of the more recent collision of NGC~7318B with these tidal structures. In this scenario, the differences between the two bridge structures are more difficult to explain; if both are the product of tidal forces and the shock from the collision of NGC~7319, why are their velocity structures different, and why does the northern bridge not show the same coherent multi-wavelength structure seen in the southern bridge? 

\subsection{Intermediate velocity structures: the shock zone}
\Hi\ located in the shock ridge at the intermediate (6200-6500\kmps, Figure\,\ref{fig::HI_a} bottom-panel) velocity range is of particular interest. This is the velocity range in which H$\alpha$ observations find the majority of the shocked, high velocity-dispersion ionized gas \citep{Arnaudova2024} associated with the shock ridge. This ionized gas seems likely to trace the material most strongly affected by the collisional shock, with its intermediate velocity indicating that is has been accelerated either from the pre-existing tidal \Hi\ filament, or out of the disk of NGC~7318B.

Part of the \Hi\ flux observed in this component by FAST \citep{Cheng2023} is now confirmed as arising from gas in the northern shock ridge coincident with the northeast arm of NGC~7318B, and in the northern part of SQ-A (Figure\,\ref{fig::HI_a} bottom-panel). However, in this velocity component we do not see the strong peak in \Hi\ column densities at the position of SQ-A found in the 6000, 6600 and 6900 components. The \Hi\ is limited to the northern half of the ridge (as is \Hi\ in the 6600 component) avoiding the regions of brightest radio continuum emission in the southern half of the ridge. This could suggest a variation in shock strength from north to south, with stronger shocks in the south effectively ionizing all of the \Hi\ in that portion of the tidal filament, while some \Hi\ survived the collision in the north. An alternative would be variation in time since collision; if the tidal \Hi\ filament were tilted out of the plane of the sky, NGC~7318B might have collided with the northern part first. In this case, the intermediate-velocity \Hi\ in the northern ridge might be gas which was initially ionized and accelerated by the collision, but has now had time to cool back into the neutral phase, while retaining its new velocity. Discriminating between these options, or other possible origin scenarios, requires information on shock strength and age. We therefore defer further discussion to paper~II, where the results of radio continuum analysis may provide insight.

\section{Summary and Conclusions}
\label{sec::summary}
In this work, we have presented MeerKAT (6 hrs on-source) \Hi\ observations of \SQ. This dataset reveals significant new structure in the atomic gas and provides a clearer view of features identified in earlier studies. With a restoring beam of $17\arcsec\times 14\arcsec$ 
and a smoothed velocity resolution of 16.5\kmps, we reach a $1\sigma$ \Hi\ column density sensitivity of $1.3\times10^{19}\,\mathrm{cm^{-2}}$. Below, we summarize our primary results:

\begin{enumerate}
\item We report the first modern interferometric detection of \Hi\ emission in NGC~7319, and the first detection of \Hi\ in NGC~7320C. In NGC~7319 the bulk of the \Hi\ is located in the western disk, with the overall spatial distribution clumpy and asymmetric, and the velocity distribution mildly offset from the stellar component. The \Hi\ appears to share a similar distribution to the previously detected molecular gas, except in the galaxy core where CO emission is at its strongest but \Hi\ emission is weak. In NGC~7320C we find $\sim$2$\times$10$^8$\Msol\ of atomic gas, with a mean velocity of 6520\kmps\ ($z$=0.0218) in agreement with past optical measurements \citep{Sulentic2001,DuartePuertas2019}. The \Hi\ emission extends $\sim$45\arcs\ (20~kpc) north-south, covering the optical disk, and has a velocity gradient suggestive of rotation. We do not find any \Hi\ connecting NGC~7320C and the \Hi\ tidal tails of \SQ.

\item We confirm the previous finding of a large mass of \Hi\ emitting gas at the group velocity ($\sim$6600\kmps) along a line of sight through the northern half of the NGC~7318A/B galaxy pair, peaking close to the position of the starburst region SQ-A. We see significant structure in this gas, with filaments extending along the optical arms of NGC~7318B, including the northern half of the shock region. We determine the location of the high-velocity ($\sim$6900\kmps) \Hi\ emission previously detected in low spatial resolution observations, finding it to be also centered on SQ-A.

\item Our new observations uncover an \Hi\ counterpart to the CO/H2/H$\alpha$ bridge linking NGC~7319 and NGC 7318B, and show the same $\sim$700\kmps\ velocity gradient along its $\sim$25~kpc length found in CO emission. This bridge is also detected in X-ray and radio continuum emission. While the origin of the bridge is unclear, its orientation and smooth velocity gradient make it difficult to explain except as a tidal feature.  

\item We detect a second \Hi\ bridge linking the northwest quadrant of NGC~7319 to SQ-A. It shows a relatively flat velocity distribution at $\sim$6600\kmps\ despite its western end being only $\sim$20~kpc north of the previously identified southern bridge. Some clumpy CO emission has been found in this northern bridge, and an X-ray filament runs along its southern edge, but there is no clear corresponding radio continuum or warm H$_2$ structure. If the bridges are tidal structures shocked by the collision of NGC~7318B with the group, this suggests that the north bridge was less affected by the shocks than its southern counterpart.

\item For the \Hi\ tidal tails extending east of the group center, the MeerKAT data shows very similar morphology to that found by the VLA. The separation between the northern and southern tails is clearer, with a definite gap in emission between the two. However, we find no evidence for distinct kinematics or separate velocity dispersion.

\item At lower velocities ($\sim$5650-6200\kmps) we find the \Hi\ emission in and around NGC~7318B is significantly more extended than was suggested by the VLA observations. \Hi\ fills much of the area of the disk with gaps only seen in the galaxy core, in the region immediately north of the core, and in a channel extending southeast across the (CO-rich, star-forming) SE arm of NGC~7318B. The velocity structure of the \Hi\ is suggestive of clockwise disk rotation, consistent with the orientation of the spiral arms of NGC~7318B. Taking into account the inclination of the galaxy, the rotational velocity of such a disk would be $\sim$175\kmps. The total gas mass would be $\sim$3$\times$10$^{11}$\Msol\ within 41~kpc radius, $\sim$53\% of the \Hi\ mass expected for NGC~7318B based on its optical luminosity. However, we note that the gas in the northern, $\sim$6000\kmps\ component appears offset from the optical redshift of the galaxy, perhaps indicating disturbance from the interaction with the higher-velocity material around SQ-A.

\item The MeerKAT data confirms the connection between the \Hi\ tails and the northern high-velocity (6600+6900\kmps) \Hi\ structures around SQ-A through the shock ridge. The southern end of the ridge splits into two branches which wrap around the base of the southern \Hi\ tail. The brightest radio continuum emission is detected in the southern bridge, where there is little or no \Hi\ emission in the group velocity range. By contrast, \Hi\ is found in this velocity range in the northern ridge, where the radio continuum emission is fainter and X-ray emission brighter. MeerKAT also localizes some of the intermediate velocity (6200-6500\kmps) \Hi\ emission previously detected by single-dish measurements, showing it to be in the northern shock ridge, north of SQ-A, and associated with the two \Hi\ bridges. This is the velocity range in which shock-excited H$\alpha$ emission dominates, implying that at least some of this \Hi\ is associated with gas strongly affected by the collision shock.

\item We measure a total \Hi\ integrated flux density of 17.2$\pm$2.0~Jy~km~s$^{-1}$ from the combined spectrum of \SQ. The corresponding atomic gas mass is M$_{\rm H\textsc{i}}$=3.5$\pm$0.4$\times$10$^{10}$\Msol\ for our adopted distance of 94~Mpc. This is slightly lower than, but consistent within uncertainties with, the single-dish measurement from FAST. Based on the spatially masked \texttt{SoFiA} maps, which are more directly comparable with prior interferometric observations, we find a lower integrated flux density of $12\pm 1\,\rm Jy\,km\,s^{-1}$ ($\geq3\sigma_{\rm rms}$), equivalent to an atomic gas mass of $M_{\rm \hi} =2.4 \pm0.2\times10^{10}\,M_\odot$. This is still a factor $\sim$1.8 higher than previous VLA estimates. These measurements suggest that MeerKAT is capturing the majority of the \Hi\ flux from the group, and therefore probably mapping all the major \Hi\ structures.

\item We detected \Hi\ emission from sixteen neighboring galaxies in the \SQ~ field, of which eleven are either previously undetected or new interferometric detections. 
\end{enumerate}

The new and expanded \Hi\ structures identified within the group have implications for our understanding of the galaxy interactions which have brought the group to its current state. Although in this work, we have focused on the \Hi\ spectral line, our MeerKAT data also provide an exceptionally sensitive window on the radio continuum emission from the group. In paper~II, we will combine the results presented here with a broad-band continuum analysis (144~MHz to 8~GHz), with the goal of providing insight into the dynamics of the group and the formation mechanism of the shock ridge.

\section*{acknowledgments}
The authors thank the staff of the MeerKAT observatory for their help with the observations presented in this work. The MeerKAT telescope is operated by the South African Radio Astronomy Observatory, which is a facility of the National Research Foundation, an agency of the Department of Science, Technology and Innovation. Some of the computations in this paper were conducted on the Smithsonian High Performance Cluster (SI/HPC), Smithsonian Institution (\url{https://doi.org/10.25572/SIHPC}). Basic research in radio astronomy at the Naval Research Laboratory is supported by 6.1 Base funding. This research has made use of the NASA/IPAC Extragalactic Database, which is funded by the National Aeronautics and Space Administration and operated by the California Institute of Technology. We acknowledge the usage of the HyperLeda database (http://leda.univ-lyon1.fr).

\facilities{MeerKAT}, {CXO} 

\software{CARACal \citep{caracal2020}, AOflagger \citep{Offringa2010}, WSClean \citep{Offringa2014}, CARTA \citep{CARTA2021}, DS9 \citep{ds9}, SoFiA \citep{Serra2015}, Astropy \citep{astropy2013, astropy2018}, APLpy \citep{aplpy}, Matplotlib \citep{matplotlib}}

\bibliographystyle{aasjournal}
\bibliography{ref}

@ARTICLE{Serra2015,
       author = {{Serra}, Paolo and {Westmeier}, Tobias and {Giese}, Nadine and {Jurek}, Russell and {Fl{\"o}er}, Lars and {Popping}, Attila and {Winkel}, Benjamin and {van der Hulst}, Thijs and {Meyer}, Martin and {Koribalski}, B{\"a}rbel S. and {Staveley-Smith}, Lister and {Courtois}, H{\'e}l{\`e}ne},
        title = "{SOFIA: a flexible source finder for 3D spectral line data}",
      journal = {\mnras},
         year = 2015,
        month = apr,
       volume = {448},
       number = {2},
        pages = {1922-1929},
          doi = {10.1093/mnras/stv079},
archivePrefix = {arXiv},
       eprint = {1501.03906},
 primaryClass = {astro-ph.IM},
       adsurl = {https://ui.adsabs.harvard.edu/abs/2015MNRAS.448.1922S}
}

@PHDTHESIS{Briggs1995,
       author = {{Briggs}, Daniel Shenon},
        title = "{High fidelity deconvolution of moderately resolved sources}",
       school = {New Mexico Institute of Mining and Technology},
         year = 1995,
        month = jan,
       adsurl = {https://ui.adsabs.harvard.edu/abs/1995PhDT.......238B}
}

@ARTICLE{Kenyon2018,
       author = {{Kenyon}, J.~S. and {Smirnov}, O.~M. and {Grobler}, T.~L. and {Perkins}, S.~J.},
        title = "{CUBICAL - fast radio interferometric calibration suite exploiting complex optimization}",
      journal = {\mnras},
         year = 2018,
        month = aug,
       volume = {478},
       number = {2},
        pages = {2399-2415},
          doi = {10.1093/mnras/sty1221},
archivePrefix = {arXiv},
       eprint = {1805.03410},
 primaryClass = {astro-ph.IM},
       adsurl = {https://ui.adsabs.harvard.edu/abs/2018MNRAS.478.2399K}
}

@ARTICLE{Offringa2014,
       author = {{Offringa}, A.~R. and {McKinley}, B. and {Hurley-Walker}, N. and
         {Briggs}, F.~H. and {Wayth}, R.~B. and {Kaplan}, D.~L. and
         {Bell}, M.~E. and {Feng}, L. and {Neben}, A.~R. and {Hughes}, J.~D. and
         {Rhee}, J. and {Murphy}, T. and {Bhat}, N.~D.~R. and {Bernardi}, G. and
         {Bowman}, J.~D. and {Cappallo}, R.~J. and {Corey}, B.~E. and {Deshpand
        e}, A.~A. and {Emrich}, D. and {Ewall-Wice}, A. and {Gaensler}, B.~M. and
         {Goeke}, R. and {Greenhill}, L.~J. and {Hazelton}, B.~J. and
         {Hindson}, L. and {Johnston-Hollitt}, M. and {Jacobs}, D.~C. and
         {Kasper}, J.~C. and {Kratzenberg}, E. and {Lenc}, E. and
         {Lonsdale}, C.~J. and {Lynch}, M.~J. and {McWhirter}, S.~R. and
         {Mitchell}, D.~A. and {Morales}, M.~F. and {Morgan}, E. and
         {Kudryavtseva}, N. and {Oberoi}, D. and {Ord}, S.~M. and {Pindor}, B. and
         {Procopio}, P. and {Prabu}, T. and {Riding}, J. and {Roshi}, D.~A. and
         {Shankar}, N. Udaya and {Srivani}, K.~S. and {Subrahmanyan}, R. and
         {Tingay}, S.~J. and {Waterson}, M. and {Webster}, R.~L. and
         {Whitney}, A.~R. and {Williams}, A. and {Williams}, C.~L.},
        title = "{WSCLEAN: an implementation of a fast, generic wide-field imager for radio astronomy}",
      journal = {\mnras},
         year = 2014,
        month = oct,
       volume = {444},
       number = {1},
        pages = {606-619},
          doi = {10.1093/mnras/stu1368},
archivePrefix = {arXiv},
       eprint = {1407.1943},
 primaryClass = {astro-ph.IM},
       adsurl = {https://ui.adsabs.harvard.edu/abs/2014MNRAS.444..606O}
}

@misc{CARTA2021,
  author       = {Angus Comrie and
                  Kuo-Song Wang and
                  Shou-Chieh Hsu and
                  Anthony Moraghan and
                  Pamela Harris and
                  Qi Pang and
                  Adrianna Pińska and
                  Cheng-Chin Chiang and
                  Tien-Hao Chang and
                  Yu-Hsuan Hwang and
                  Hengtai Jan and
                  Ming-Yi Lin and
                  Rob Simmonds},
  title        = {{CARTA: The Cube Analysis and Rendering Tool for Astronomy}},
  month        = jun,
  year         = 2021,
  publisher    = {Zenodo},
  version      = {2.0.0},
  doi          = {10.5281/zenodo.4905459},
  url          = {https://doi.org/10.5281/zenodo.4905459}
}

@inproceedings{ds9,
  author    = {{Joye}, W.~A. and {Mandel}, E.},
  title     = {{New Features of SAOImage DS9}},
  booktitle = {Astronomical Data Analysis Software and Systems XII},
  year      = 2003,
  editor    = {{Payne}, H.~E. and {Jedrzejewski}, R.~I. and {Hook}, R.~N.},
  series    = {Astronomical Society of the Pacific Conference Series},
  volume    = {295},
  month     = jan,
  pages     = {489},
  adsurl    = {https://ui.adsabs.harvard.edu/abs/2003ASPC..295..489J}
}

@ARTICLE{Astropy2013,
   author = {{Astropy Collaboration} and {Robitaille}, T.~P. and {Tollerud}, E.~J. and 
	{Greenfield}, P. and {Droettboom}, M. and {Bray}, E. and {Aldcroft}, T. and 
	{Davis}, M. and {Ginsburg}, A. and {Price-Whelan}, A.~M. and 
	{Kerzendorf}, W.~E. and {Conley}, A. and {Crighton}, N. and 
	{Barbary}, K. and {Muna}, D. and {Ferguson}, H. and {Grollier}, F. and 
	{Parikh}, M.~M. and {Nair}, P.~H. and {Unther}, H.~M. and {Deil}, C. and 
	{Woillez}, J. and {Conseil}, S. and {Kramer}, R. and {Turner}, J.~E.~H. and 
	{Singer}, L. and {Fox}, R. and {Weaver}, B.~A. and {Zabalza}, V. and 
	{Edwards}, Z.~I. and {Azalee Bostroem}, K. and {Burke}, D.~J. and 
	{Casey}, A.~R. and {Crawford}, S.~M. and {Dencheva}, N. and 
	{Ely}, J. and {Jenness}, T. and {Labrie}, K. and {Lim}, P.~L. and 
	{Pierfederici}, F. and {Pontzen}, A. and {Ptak}, A. and {Refsdal}, B. and 
	{Servillat}, M. and {Streicher}, O.},
    title = "{Astropy: A community Python package for astronomy}",
  journal = {\aap},
archivePrefix = "arXiv",
   eprint = {1307.6212},
 primaryClass = "astro-ph.IM",
     year = 2013,
    month = oct,
   volume = 558,
      eid = {A33},
    pages = {A33},
      doi = {10.1051/0004-6361/201322068},
   adsurl = {http://adsabs.harvard.edu/abs/2013A%26A...558A..33A}
}

@ARTICLE{astropy2018,
       author = {{Astropy Collaboration} and {Price-Whelan}, A.~M. and {Sip{\H{o}}cz}, B.~M. and {G{\"u}nther}, H.~M. and {Lim}, P.~L. and {Crawford}, S.~M. and {Conseil}, S. and {Shupe}, D.~L. and {Craig}, M.~W. and {Dencheva}, N. and {Ginsburg}, A. and {VanderPlas}, J.~T. and {Bradley}, L.~D. and {P{\'e}rez-Su{\'a}rez}, D. and {de Val-Borro}, M. and {Aldcroft}, T.~L. and {Cruz}, K.~L. and {Robitaille}, T.~P. and {Tollerud}, E.~J. and {Ardelean}, C. and {Babej}, T. and {Bach}, Y.~P. and {Bachetti}, M. and {Bakanov}, A.~V. and {Bamford}, S.~P. and {Barentsen}, G. and {Barmby}, P. and {Baumbach}, A. and {Berry}, K.~L. and {Biscani}, F. and {Boquien}, M. and {Bostroem}, K.~A. and {Bouma}, L.~G. and {Brammer}, G.~B. and {Bray}, E.~M. and {Breytenbach}, H. and {Buddelmeijer}, H. and {Burke}, D.~J. and {Calderone}, G. and {Cano Rodr{\'\i}guez}, J.~L. and {Cara}, M. and {Cardoso}, J.~V.~M. and {Cheedella}, S. and {Copin}, Y. and {Corrales}, L. and {Crichton}, D. and {D'Avella}, D. and {Deil}, C. and {Depagne}, {\'E}. and {Dietrich}, J.~P. and {Donath}, A. and {Droettboom}, M. and {Earl}, N. and {Erben}, T. and {Fabbro}, S. and {Ferreira}, L.~A. and {Finethy}, T. and {Fox}, R.~T. and {Garrison}, L.~H. and {Gibbons}, S.~L.~J. and {Goldstein}, D.~A. and {Gommers}, R. and {Greco}, J.~P. and {Greenfield}, P. and {Groener}, A.~M. and {Grollier}, F. and {Hagen}, A. and {Hirst}, P. and {Homeier}, D. and {Horton}, A.~J. and {Hosseinzadeh}, G. and {Hu}, L. and {Hunkeler}, J.~S. and {Ivezi{\'c}}, {\v{Z}}. and {Jain}, A. and {Jenness}, T. and {Kanarek}, G. and {Kendrew}, S. and {Kern}, N.~S. and {Kerzendorf}, W.~E. and {Khvalko}, A. and {King}, J. and {Kirkby}, D. and {Kulkarni}, A.~M. and {Kumar}, A. and {Lee}, A. and {Lenz}, D. and {Littlefair}, S.~P. and {Ma}, Z. and {Macleod}, D.~M. and {Mastropietro}, M. and {McCully}, C. and {Montagnac}, S. and {Morris}, B.~M. and {Mueller}, M. and {Mumford}, S.~J. and {Muna}, D. and {Murphy}, N.~A. and {Nelson}, S. and {Nguyen}, G.~H. and {Ninan}, J.~P. and {N{\"o}the}, M. and {Ogaz}, S. and {Oh}, S. and {Parejko}, J.~K. and {Parley}, N. and {Pascual}, S. and {Patil}, R. and {Patil}, A.~A. and {Plunkett}, A.~L. and {Prochaska}, J.~X. and {Rastogi}, T. and {Reddy Janga}, V. and {Sabater}, J. and {Sakurikar}, P. and {Seifert}, M. and {Sherbert}, L.~E. and {Sherwood-Taylor}, H. and {Shih}, A.~Y. and {Sick}, J. and {Silbiger}, M.~T. and {Singanamalla}, S. and {Singer}, L.~P. and {Sladen}, P.~H. and {Sooley}, K.~A. and {Sornarajah}, S. and {Streicher}, O. and {Teuben}, P. and {Thomas}, S.~W. and {Tremblay}, G.~R. and {Turner}, J.~E.~H. and {Terr{\'o}n}, V. and {van Kerkwijk}, M.~H. and {de la Vega}, A. and {Watkins}, L.~L. and {Weaver}, B.~A. and {Whitmore}, J.~B. and {Woillez}, J. and {Zabalza}, V. and {Astropy Contributors}},
        title = "{The Astropy Project: Building an Open-science Project and Status of the v2.0 Core Package}",
      journal = {\aj},
         year = 2018,
        month = sep,
       volume = {156},
       number = {3},
          eid = {123},
        pages = {123},
          doi = {10.3847/1538-3881/aabc4f},
archivePrefix = {arXiv},
       eprint = {1801.02634},
 primaryClass = {astro-ph.IM},
       adsurl = {https://ui.adsabs.harvard.edu/abs/2018AJ....156..123A}
}

@ARTICLE{VerdesMontenegro2001,
       author = {{Verdes-Montenegro}, L. and {Yun}, M.~S. and {Williams}, B.~A. and {Huchtmeier}, W.~K. and {Del Olmo}, A. and {Perea}, J.},
        title = "{Where is the neutral atomic gas in Hickson groups?}",
      journal = {\aap},
         year = 2001,
        month = oct,
       volume = {377},
        pages = {812-826},
          doi = {10.1051/0004-6361:20011127},
archivePrefix = {arXiv},
       eprint = {astro-ph/0108223},
 primaryClass = {astro-ph},
       adsurl = {https://ui.adsabs.harvard.edu/abs/2001A&A...377..812V}
}

@ARTICLE{Xu2003,
       author = {{Xu}, C.~K. and {Lu}, N. and {Condon}, J.~J. and {Dopita}, M. and {Tuffs}, R.~J.},
        title = "{Physical Conditions and Star Formation Activity in the Intragroup Medium of Stephan's Quintet}",
      journal = {\apj},
         year = 2003,
        month = oct,
       volume = {595},
       number = {2},
        pages = {665-684},
          doi = {10.1086/377445}}

@ARTICLE{Allen1972,
       author = {{Allen}, Ronald J. and {Hartsuiker}, Jacob W.},
        title = "{Radio Continuum Emission at 21 cm near Stephan's Quintet}",
      journal = {\nat},
         year = 1972,
        month = oct,
       volume = {239},
       number = {5371},
        pages = {324-325},
          doi = {10.1038/239324a0}}

@ARTICLE{vanderHulst1981,
       author = {{van der Hulst}, J.~M. and {Rots}, A.~H.},
        title = "{VLA observations of the radio continuum emission from Stephan's Quintet.}",
      journal = {\aj},
         year = 1981,
        month = dec,
       volume = {86},
        pages = {1775-1780},
          doi = {10.1086/113060}}

@ARTICLE{Williams2002,
       author = {{Williams}, B.~A. and {Yun}, Min S. and {Verdes-Montenegro}, L.},
        title = "{The VLA H I Observations of Stephan's Quintet (HCG 92)}",
      journal = {\aj},
         year = 2002,
        month = may,
       volume = {123},
       number = {5},
        pages = {2417-2437},
          doi = {10.1086/339839}}

@ARTICLE{Shostak1984,
       author = {{Shostak}, G.~S. and {Sullivan}, III, W.~T. and {Allen}, R.~J.},
        title = "{H I synthesis observations of the high-redshift galaxies in Stephan'sQuintet.}",
      journal = {\aap},
         year = 1984,
        month = oct,
       volume = {139},
        pages = {15-24}}

@ARTICLE{Arnaudova2024,
       author = {{Arnaudova}, M.~I. and {Das}, S. and {Smith}, D.~J.~B. and {Hardcastle}, M.~J. and {Hatch}, N. and {Trager}, S.~C. and {Smith}, R.~J. and {Drake}, A.~B. and {McGarry}, J.~C. and {Shenoy}, S. and {Stott}, J.~P. and {Knapen}, J.~H. and {Hess}, K.~M. and {Duncan}, K.~J. and {Gloudemans}, A. and {Best}, P.~N. and {Garc{\'\i}a-Benito}, R. and {Kondapally}, R. and {Balcells}, M. and {Couto}, G.~S. and {Abrams}, D.~C. and {Aguado}, D. and {Aguerri}, J.~A.~L. and {Barrena}, R. and {Benn}, C.~R. and {Bensby}, T. and {Berlanas}, S.~R. and {Bettoni}, D. and {Cano-Infantes}, D. and {Carrera}, R. and {Concepci{\'o}n}, P.~J. and {Dalton}, G.~B. and {D'Ago}, G. and {Dee}, K. and {Dom{\'\i}nguez-Palmero}, L. and {Drew}, J.~E. and {Escott}, E.~L. and {Fari{\~n}a}, C. and {Fossati}, M. and {Fumagalli}, M. and {Gafton}, E. and {Gribbin}, F.~J. and {Hughes}, S. and {Iovino}, A. and {Jin}, S. and {Lewis}, I.~J. and {Longhetti}, M. and {M{\'e}ndez-Abreu}, J. and {Mercurio}, A. and {Molaeinezhad}, A. and {Molinari}, E. and {Mongui{\'o}}, M. and {Murphy}, D.~N.~A. and {Pic{\'o}}, S. and {Pieri}, M.~M. and {Ridings}, A.~W. and {Romero-G{\'o}mez}, M. and {Schallig}, E. and {Shimwell}, T.~W. and {Skvar{\v{c}}}, J. and {Stuik}, R. and {Vallenari}, A. and {van der Hulst}, J.~M. and {Walton}, N.~A. and {Worley}, C.~C.},
        title = "{WEAVE First Light Observations: Origin and Dynamics of the Shock Front in Stephan's Quintet}",
      journal = {\mnras},
         year = 2024,
        month = dec,
       volume = {535},
       number = {3},
        pages = {2269-2290},
          doi = {10.1093/mnras/stae2235}}

@ARTICLE{Xanthopoulos2004,
       author = {{Xanthopoulos}, E. and {Muxlow}, T.~W.~B. and {Thomasson}, P. and {Garrington}, S.~T.},
        title = "{MERLIN observations of Stephan's Quintet}",
      journal = {\mnras},
         year = 2004,
        month = oct,
       volume = {353},
       number = {4},
        pages = {1117-1125},
          doi = {10.1111/j.1365-2966.2004.08133.x}}

@ARTICLE{Konstantopoulos2010,
   author = {{Konstantopoulos}, I.~S. and {Gallagher}, S.~C. and {Fedotov}, K. and {Durrell}, P.~R. and {Heiderman}, A. and {Elmegreen}, D.~M. and {Charlton}, J.~C. and {Hibbard}, J.~E. and {Tzanavaris}, P. and {Chandar}, R. and {Johnson}, K.~E. and {et al.}},
    title = "{Galaxy Evolution in a Complex Environment: A Multi-wavelength Study of HCG 7}",
  journal = {\apj},
     year = 2010,
    month = nov,
   volume = 723,
    pages = {197},
      doi = {10.1088/0004-637X/723/1/197}}

@ARTICLE{OSullivan2009,
   author = {{O'Sullivan}, E. and {Giacintucci}, S. and {Vrtilek}, J.~M. and {Raychaudhury}, S. and {David}, L.~P.},
    title = "{A Chandra X-ray View of Stephan's Quintet: Shocks and Star Formation}",
  journal = {\apj},
archivePrefix = "arXiv",
   eprint = {0812.0383},
     year = 2009,
    month = aug,
   volume = 701,
    pages = {1560},
      doi = {10.1088/0004-637X/701/2/1560}}

@ARTICLE{Sulentic2001,
   author = {{Sulentic}, J.~W. and {Rosado}, M. and {Dultzin-Hacyan}, D. and {Verdes-Montenegro}, L. and {Trinchieri}, G. and {Xu}, C. and {Pietsch}, W.},
    title = "{A Multiwavelength Study of Stephan's Quintet}",
  journal = {\aj},
   eprint = {arXiv:astro-ph/0111155},
     year = 2001,
    month = dec,
   volume = 122,
    pages = {2993},
      doi = {10.1086/324455}}

@ARTICLE{Appleton2013,
       author = {{Appleton}, P.~N. and {Guillard}, P. and {Boulanger}, F. and {Cluver}, M.~E. and {Ogle}, P. and {Falgarone}, E. and {Pineau des For{\^e}ts}, G. and {O'Sullivan}, E. and {Duc}, P. -A. and {Gallagher}, S. and {Gao}, Y. and {Jarrett}, T. and {Konstantopoulos}, I. and {Lisenfeld}, U. and {Lord}, S. and {Lu}, N. and {Peterson}, B.~W. and {Struck}, C. and {Sturm}, E. and {Tuffs}, R. and {Valchanov}, I. and {van der Werf}, P. and {Xu}, K.~C.},
        title = "{Shock-enhanced C$^{+}$ Emission and the Detection of H$_{2}$O from the Stephan's Quintet Group-wide Shock Using Herschel}",
      journal = {\apj},
         year = 2013,
        month = nov,
       volume = {777},
       number = {1},
          eid = {66},
        pages = {66},
          doi = {10.1088/0004-637X/777/1/66}}

@ARTICLE{Appleton2017,
       author = {{Appleton}, P.~N. and {Guillard}, P. and {Togi}, A. and {Alatalo}, K. and {Boulanger}, F. and {Cluver}, M. and {Pineau des For{\^e}ts}, G. and {Lisenfeld}, U. and {Ogle}, P. and {Xu}, C.~K.},
        title = "{Powerful H$_{2}$ Line Cooling in Stephan{\textquoteright}s Quintet. II. Group-wide Gas and Shock Modeling of the Warm H$_{2}$ and a Comparison with [C II] 157.7 {\ensuremath{\mu}}m Emission and Kinematics}",
      journal = {\apj},
         year = 2017,
        month = feb,
       volume = {836},
       number = {1},
          eid = {76},
        pages = {76},
          doi = {10.3847/1538-4357/836/1/76}}

@ARTICLE{DuartePuertas2019,
       author = {{Duarte Puertas}, S. and {Iglesias-P{\'a}ramo}, J. and {Vilchez}, J.~M. and {Drissen}, L. and {Kehrig}, C. and {Martin}, T.},
        title = "{Searching for intergalactic star forming regions in Stephan's Quintet with SITELLE. I. Ionised gas structures and kinematics}",
      journal = {\aap},
         year = 2019,
        month = sep,
       volume = {629},
          eid = {A102},
        pages = {A102},
          doi = {10.1051/0004-6361/201935686}}

@ARTICLE{Emonts2025,
       author = {{Emonts}, B.~H.~C. and {Appleton}, P.~N. and {Lisenfeld}, U. and {Guillard}, P. and {Xu}, C.~K. and {Reach}, W.~T. and {Barcos-Mu{\~n}oz}, L. and {Labiano}, A. and {Ogle}, P.~M. and {O'Sullivan}, E. and {Togi}, A. and {Gallagher}, S.~C. and {Aromal}, P. and {Duc}, P. -A. and {Alatalo}, K. and {Boulanger}, F. and {D{\'\i}az-Santos}, T. and {Helou}, G.},
        title = "{Bird's-eye View of Molecular Gas across Stephan's Quintet Galaxy Group and Intragroup Medium}",
      journal = {\apj},
         year = 2025,
        month = jan,
       volume = {978},
       number = {1},
          eid = {111},
        pages = {111},
          doi = {10.3847/1538-4357/ad957c}}

@ARTICLE{Appleton2023,
       author = {{Appleton}, P.~N. and {Guillard}, P. and {Emonts}, Bjorn and {Boulanger}, Francois and {Togi}, Aditya and {Reach}, William T. and {Alatalo}, Katherine and {Cluver}, M. and {Diaz Santos}, T. and {Duc}, P. -A. and {Gallagher}, S. and {Ogle}, P. and {O'Sullivan}, E. and {Voggel}, K. and {Xu}, C.~K.},
        title = "{Multiphase Gas Interactions on Subarcsec Scales in the Shocked Intergalactic Medium of Stephan's Quintet with JWST and ALMA}",
      journal = {\apj},
         year = 2023,
        month = jul,
       volume = {951},
       number = {2},
          eid = {104},
        pages = {104},
          doi = {10.3847/1538-4357/accc2a}}

@ARTICLE{Guillard2012,
       author = {{Guillard}, P. and {Boulanger}, F. and {Pineau des For{\^e}ts}, G. and {Falgarone}, E. and {Gusdorf}, A. and {Cluver}, M.~E. and {Appleton}, P.~N. and {Lisenfeld}, U. and {Duc}, P. -A. and {Ogle}, P.~M. and {Xu}, C.~K.},
        title = "{Turbulent Molecular Gas and Star Formation in the Shocked Intergalactic Medium of Stephan's Quintet}",
      journal = {\apj},
         year = 2012,
        month = apr,
       volume = {749},
       number = {2},
          eid = {158},
        pages = {158},
          doi = {10.1088/0004-637X/749/2/158}}

@ARTICLE{Xu2022,
       author = {{Xu}, C.~K. and {Cheng}, C. and {Appleton}, P.~N. and {Duc}, P. -A. and {Gao}, Y. and {Tang}, N. -Y. and {Yun}, M. and {Dai}, Y.~S. and {Huang}, J. -S. and {Lisenfeld}, U. and {Renaud}, F.},
        title = "{A 0.6 Mpc H I structure associated with Stephan's Quintet}",
      journal = {\nat},
         year = 2022,
        month = oct,
       volume = {610},
       number = {7932},
        pages = {461-466},
          doi = {10.1038/s41586-022-05206-x}
}

@ARTICLE{Cheng2023,
       author = {{Cheng}, Cheng and {Xu}, Cong Kevin and {Appleton}, P.~N. and {Duc}, P. -A. and {Tang}, N. -Y. and {Dai}, Y. -S. and {Huang}, J. -S. and {Lisenfeld}, U. and {Renaud}, F. and {He}, Chuan and {Feng}, Hai-Cheng},
        title = "{Deep H I Mapping of Stephan's Quintet and Its Neighborhood}",
      journal = {\apj},
         year = 2023,
        month = sep,
       volume = {954},
       number = {1},
          eid = {74},
        pages = {74},
          doi = {10.3847/1538-4357/ace03e}
}

@ARTICLE{Jones2023,
       author = {{Jones}, M.~G. and {Verdes-Montenegro}, L. and {Moldon}, J. and {Damas Segovia}, A. and {Borthakur}, S. and {Luna}, S. and {Yun}, M. and {del Olmo}, A. and {Perea}, J. and {Cannon}, J. and {Lopez Gutierrez}, D. and {Cluver}, M. and {Garrido}, J. and {Sanchez}, S.},
        title = "{Disturbed, diffuse, or just missing? A global study of the H I content of Hickson compact groups}",
      journal = {\aap},
         year = 2023,
        month = feb,
       volume = {670},
          eid = {A21},
        pages = {A21},
          doi = {10.1051/0004-6361/202244622}
}

@ARTICLE{Borthakur2010,
       author = {{Borthakur}, Sanchayeeta and {Yun}, Min Su and {Verdes-Montenegro}, Lourdes},
        title = "{Detection of Diffuse Neutral Intragroup Medium in Hickson Compact Groups}",
      journal = {\apj},
         year = 2010,
        month = feb,
       volume = {710},
       number = {1},
        pages = {385-407},
          doi = {10.1088/0004-637X/710/1/385}
}

@ARTICLE{Borthakur2015,
       author = {{Borthakur}, Sanchayeeta and {Yun}, Min Su and {Verdes-Montenegro}, Lourdes and {Heckman}, Timothy M. and {Zhu}, Guangtun and {Braatz}, James A.},
        title = "{Distribution of Faint Atomic Gas in Hickson Compact Groups}",
      journal = {\apj},
         year = 2015,
        month = oct,
       volume = {812},
       number = {1},
          eid = {78},
        pages = {78},
          doi = {10.1088/0004-637X/812/1/78}
}

@ARTICLE{Balkowski1973,
       author = {{Balkowski}, C. and {Bottinelli}, L. and {Chamaraux}, P. and {Gouguenheim}, L. and {Heidmann}, J.},
        title = "{Distance of two galaxies in Stephan's quintet and possible non-velocity redshifts.}",
      journal = {\aap},
         year = 1973,
        month = jun,
       volume = {25},
        pages = {319-324}
}

@ARTICLE{Shostak1974,
       author = {{Shostak}, G.~S.},
        title = "{H I emission from Stephan's Quintet.}",
      journal = {\apj},
         year = 1974,
        month = jan,
       volume = {187},
        pages = {19-23},
          doi = {10.1086/152584}
}

@ARTICLE{Allen1980,
       author = {{Allen}, R.~J. and {Sullivan}, III, W.~T.},
        title = "{The low and high redshift neutral hydrogen associated with Stephan's Quintet.}",
      journal = {\aap},
         year = 1980,
        month = apr,
       volume = {84},
        pages = {181-190}
}

@article{matplotlib,
  author    = {Hunter, J. D.},
  title     = {Matplotlib: A 2D graphics environment},
  journal   = {Computing in Science \& Engineering},
  volume    = {9},
  number    = {3},
  pages     = {90--95},
  publisher = {IEEE COMPUTER SOC},
  doi       = {10.1109/MCSE.2007.55},
  year      = 2007
}

@misc{aplpy,
  author        = {{Robitaille}, Thomas and {Bressert}, Eli},
  title         = {{APLpy: Astronomical Plotting Library in Python}},
  year          = 2012,
  month         = aug,
  eid           = {ascl:1208.017},
  pages         = {ascl:1208.017},
  archiveprefix = {ascl},
  eprint        = {1208.017},
  adsurl        = {https://ui.adsabs.harvard.edu/abs/2012ascl.soft08017R}
}

@ARTICLE{Offringa2010,
       author = {{Offringa}, A.~R. and {de Bruyn}, A.~G. and {Biehl}, M. and
         {Zaroubi}, S. and {Bernardi}, G. and {Pandey}, V.~N.},
        title = "{Post-correlation radio frequency interference classification methods}",
      journal = {\mnras},
         year = "2010",
        month = "Jun",
       volume = {405},
       number = {1},
        pages = {155-167},
          doi = {10.1111/j.1365-2966.2010.16471.x},
archivePrefix = {arXiv},
       eprint = {1002.1957},
 primaryClass = {astro-ph.IM},
       adsurl = {https://ui.adsabs.harvard.edu/abs/2010MNRAS.405..155O}
}

@MISC{caracal2020,
       author = {{J{\'o}zsa}, Gyula I.~G. and {White}, Sarah V. and {Thorat}, Kshitij and {Smirnov}, Oleg M. and {Serra}, Paolo and {Ramatsoku}, Mpati and {Ramaila}, Athanaseus J.~T. and {Perkins}, Simon J. and {Moln{\'a}r}, D{\'a}niel Cs. and {Makhathini}, Sphesihle and {Maccagni}, Filippo M. and {Kleiner}, Dane and {Kamphuis}, Peter and {Hugo}, Benjamin V. and {de Blok}, W.~J.~G. and {Andati}, Lexy A.~L.},
        title = "{CARACal: Containerized Automated Radio Astronomy Calibration pipeline}",
 howpublished = {Astrophysics Source Code Library, record ascl:2006.014},
         year = 2020,
        month = jun,
          eid = {ascl:2006.014},
        pages = {ascl:2006.014},
archivePrefix = {ascl},
       eprint = {2006.014},
       adsurl = {https://ui.adsabs.harvard.edu/abs/2020ascl.soft06014J}
}

@ARTICLE{Verdes-Montenegro2001,
       author = {{Verdes-Montenegro}, L. and {Yun}, M.~S. and {Williams}, B.~A. and {Huchtmeier}, W.~K. and {Del Olmo}, A. and {Perea}, J.},
        title = "{Where is the neutral atomic gas in Hickson groups?}",
      journal = {\aap},
         year = 2001,
        month = oct,
       volume = {377},
        pages = {812-826},
          doi = {10.1051/0004-6361:20011127},
archivePrefix = {arXiv},
       eprint = {astro-ph/0108223},
 primaryClass = {astro-ph},
       adsurl = {https://ui.adsabs.harvard.edu/abs/2001A&A...377..812V}
}

@ARTICLE{Trinchieri2003,
       author = {{Trinchieri}, G. and {Sulentic}, J. and {Breitschwerdt}, D. and {Pietsch}, W.},
        title = "{Stephan's Quintet: The X-ray anatomy of a multiple galaxy collision}",
      journal = {\aap},
         year = 2003,
        month = apr,
       volume = {401},
        pages = {173-183},
          doi = {10.1051/0004-6361:20030108}}

@ARTICLE{Trinchierietal05,
   author = {{Trinchieri}, G. and {Sulentic}, J. and {Pietsch}, W. and {Breitschwerdt}, D.},
    title = "{Stephan's Quintet with XMM-Newton}",
  journal = {\aap},
   eprint = {arXiv:astro-ph/0506761},
     year = 2005,
    month = dec,
   volume = 444,
    pages = {697},
      doi = {10.1051/0004-6361:20052910}}

@ARTICLE{Stephan1877,
   author = {{Stephan}, M.},
    title = "{Nebul{\ae} (new) discovered and observed at the observatory of Marseilles, 1876 and 1877, M. Stephan}",
  journal = {\mnras},
     year = 1877,
    month = apr,
   volume = 37,
    pages = {334}}

@article{Hickson82,
        author = "Hickson, P.",
        title = "Systematic properties of compact groups of galaxies",
        journal = {\apj},
        year = "1982",
        volume = "255",
        pages = "382"}

@ARTICLE{Hicksonetal92,
   author = {{Hickson}, P. and {Mendes de Oliveira}, C. and {Huchra}, J.~P. and {Palumbo}, G.~G.},
    title = "{Dynamical properties of compact groups of galaxies}",
  journal = {\apj},
     year = 1992,
    month = nov,
   volume = 399,
    pages = {353},
      doi = {10.1086/171932}}

@ARTICLE{Xuetal25,
       author = {{Xu}, C.~K. and {Cheng}, C. and {Yun}, M.~S. and {Appleton}, P.~N. and {Emonts}, B.~H.~C. and {Braine}, J. and {Gallagher}, S.~C. and {Guillard}, P. and {Lisenfeld}, U. and {O'Sullivan}, E. and {Renaud}, F. and {Aromal}, P. and {Duc}, P.-A. and {Labiano}, A. and {Togi}, A.},
        title = "{SQ-A: A Collision-triggered Starburst in the Intragroup Medium of Stephan's Quintet}",
      journal = {\apj},
         year = 2025,
        month = oct,
       volume = {991},
       number = {2},
          eid = {197},
        pages = {197},
          doi = {10.3847/1538-4357/adfbf6}}

@ARTICLE{Molesetal97,
   author = {{Moles}, M. and {Sulentic}, J.~W. and {Marquez}, I.},
    title = "{The Dynamical Status of Stephan's Quintet}",
  journal = {\apj},
     year = 1997,
    month = aug,
   volume = 485,
    pages = {L69},
      doi = {10.1086/310817}}

@ARTICLE{Molesetal98,
   author = {{Moles}, M. and {Marquez}, I. and {Sulentic}, J.~W.},
    title = "{The observational status of Stephan's Quintet}",
  journal = {\aap},
     year = 1998,
    month = jun,
   volume = 334,
    pages = {473}}

@ARTICLE{Makarovetal14,
       author = {{Makarov}, Dmitry and {Prugniel}, Philippe and {Terekhova}, Nataliya and {Courtois}, H{\'e}l{\`e}ne and {Vauglin}, Isabelle},
        title = "{HyperLEDA. III. The catalogue of extragalactic distances}",
      journal = {\aap},
         year = 2014,
        month = oct,
       volume = {570},
          eid = {A13},
        pages = {A13},
          doi = {10.1051/0004-6361/201423496}}

@ARTICLE{Nishiuraetal00,
       author = {{Nishiura}, Shingo and {Shimada}, Masashi and {Ohyama}, Youichi and {Murayama}, Takashi and {Taniguchi}, Yoshiaki},
        title = "{A Dynamical Study of Galaxies in the Hickson Compact Groups}",
      journal = {\aj},
         year = 2000,
        month = oct,
       volume = {120},
       number = {4},
        pages = {1691-1712},
          doi = {10.1086/301561}}

@ARTICLE{Xuetal99,
       author = {{Xu}, Cong and {Sulentic}, Jack W. and {Tuffs}, Richard},
        title = "{Starburst in the Intragroup Medium of Stephan's Quintet}",
      journal = {\apj},
         year = 1999,
        month = feb,
       volume = {512},
       number = {1},
        pages = {178-183},
          doi = {10.1086/306771}}

@ARTICLE{Lisenfeldetal02,
       author = {{Lisenfeld}, U. and {Braine}, J. and {Duc}, P.-A. and {Leon}, S. and {Charmandaris}, V. and {Brinks}, E.},
        title = "{Abundant molecular gas in the intergalactic medium of Stephan's Quintet}",
      journal = {\aap},
         year = 2002,
        month = nov,
       volume = {394},
        pages = {823-833},
          doi = {10.1051/0004-6361:20021232}
}

@ARTICLE{Lisenfeldetal04,
       author = {{Lisenfeld}, U. and {Braine}, J. and {Duc}, P.-A. and {Brinks}, E. and {Charmandaris}, V. and {Leon}, S.},
        title = "{Molecular and ionized gas in the tidal tail in Stephan's Quintet}",
      journal = {\aap},
         year = 2004,
        month = nov,
       volume = {426},
        pages = {471-479},
          doi = {10.1051/0004-6361:20041330}}

@ARTICLE{Maedaetal2025,
       author = {{Maeda}, Fumiya and {Komugi}, Shinya and {Muraoka}, Kazuyuki and {Yamamoto}, Misaki and {Egusa}, Fumi and {Ohta}, Kouji and {Asada}, Yoshihisa and {Habe}, Asao and {Hatsukade}, Bunyo and {Kaneko}, Hiroyuki and {Kobayashi}, Masato I.~N. and {Kohno}, Kotaro and {Konishi}, Ayu and {Matsusaka}, Ren and {Morokuma-Matsui}, Kana and {Nakanishi}, Kouichiro and {Tosaki}, Tomoka and {Tsujita}, Akiyoshi},
        title = "{Spatially and Dynamically Extended Molecular Gas in Stephan's Quintet Revealed by ALMA CO(1{\textendash}0) Total Power Mapping}",
      journal = {\apj},
         year = 2025,
        month = sep,
       volume = {990},
       number = {2},
          eid = {221},
        pages = {221},
          doi = {10.3847/1538-4357/adfc56}}

@ARTICLE{Ducetal18,
       author = {{Duc}, Pierre-Alain and {Cuillandre}, Jean-Charles and {Renaud}, Florent},
        title = "{Revisiting Stephan's Quintet with deep optical images}",
      journal = {\mnras},
         year = 2018,
        month = mar,
       volume = {475},
       number = {1},
        pages = {L40-L44},
          doi = {10.1093/mnrasl/sly004}}

@ARTICLE{Yttergrenetal21,
       author = {{Yttergren}, M. and {Misquitta}, P. and {S{\'a}nchez-Monge}, {\'A}. and {Valencia-S}, M. and {Eckart}, A. and {Zensus}, A. and {Peitl-Thiesen}, T.},
        title = "{Gas and stellar dynamics in Stephan's Quintet. Mapping the kinematics in a closely interacting compact galaxy group}",
      journal = {\aap},
         year = 2021,
        month = dec,
       volume = {656},
          eid = {A83},
        pages = {A83},
          doi = {10.1051/0004-6361/202040188}
}

@ARTICLE{SolomonVandenbout05,
   author = {{Solomon}, P.~M. and {Vanden Bout}, P.~A.},
    title = "{Molecular Gas at High Redshift}",
  journal = {\araa},
     year = 2005,
    month = sep,
   volume = 43,
    pages = {677},
      doi = {10.1146/annurev.astro.43.051804.102221}
}

@ARTICLE{Pontoppidanetal22,
       author = {{Pontoppidan}, Klaus M. and {Barrientes}, Jaclyn and {Blome}, Claire and {Braun}, Hannah and {Brown}, Matthew and {Carruthers}, Margaret and {Coe}, Dan and {DePasquale}, Joseph and {Espinoza}, N{\'e}stor and {Marin}, Macarena Garcia and {Gordon}, Karl D. and {Henry}, Alaina and {Hustak}, Leah and {James}, Andi and {Jenkins}, Ann and {Koekemoer}, Anton M. and {LaMassa}, Stephanie and {Law}, David and {Lockwood}, Alexandra and {Moro-Martin}, Amaya and {Mullally}, Susan E. and {Pagan}, Alyssa and {Player}, Dani and {Proffitt}, Charles and {Pulliam}, Christine and {Ramsay}, Leah and {Ravindranath}, Swara and {Reid}, Neill and {Robberto}, Massimo and {Sabbi}, Elena and {Ubeda}, Leonardo and {Balogh}, Michael and {Flanagan}, Kathryn and {Gardner}, Jonathan and {Hasan}, Hashima and {Meinke}, Bonnie and {Nota}, Antonella},
        title = "{The JWST Early Release Observations}",
      journal = {\apjl},
         year = 2022,
        month = sep,
       volume = {936},
       number = {1},
          eid = {L14},
        pages = {L14},
          doi = {10.3847/2041-8213/ac8a4e}
}

@Misc{PontoppidanGordon22,
  author = {{Pontoppidan}, K. M. and {Gordon}, K.},
  title = "{Data from the JWST/ERO progarm}",
  publisher = "STScI/MAST",
  month = oct,
  year = 2022,
  doi = {10.1709/67ft-nb86}
}

@ARTICLE{Guillardetal22,
       author = {{Guillard}, P. and {Appleton}, P.~N. and {Boulanger}, F. and {Shull}, J.~M. and {Lehnert}, M.~D. and {Pineau des Forets}, G. and {Falgarone}, E. and {Cluver}, M.~E. and {Xu}, C.~K. and {Gallagher}, S.~C. and {Duc}, P.~A.},
        title = "{Extremely Broad Ly{\ensuremath{\alpha}} Line Emission from the Molecular Intragroup Medium in Stephan's Quintet: Evidence for a Turbulent Cascade in a Highly Clumpy Multiphase Medium?}",
      journal = {\apj},
         year = 2022,
        month = jan,
       volume = {925},
       number = {1},
          eid = {63},
        pages = {63},
          doi = {10.3847/1538-4357/ac313f}
}

@ARTICLE{Cluveretal10,
   author = {{Cluver}, M.~E. and {Appleton}, P.~N. and {Boulanger}, F. and {Guillard}, P. and {Ogle}, P. and {Duc}, P.-A. and {Lu}, N. and {Rasmussen}, J. and {Reach}, W.~T. and {Smith}, J.~D. and {Tuffs}, R. and {Xu}, C.~K. and {Yun}, M.~S.},
    title = "{Powerful H$_{2}$ Line Cooling in Stephan's Quintet. I. Mapping the Significant Cooling Pathways in Group-wide Shocks}",
  journal = {\apj},
     year = 2010,
    month = feb,
   volume = 710,
    pages = {248},
      doi = {10.1088/0004-637X/710/1/248}
}

@ARTICLE{Aokietal96,
       author = {{Aoki}, Kentaro and {Ohtani}, Hiroshi and {Yoshida}, Michitosh and {Kosugi}, George},
        title = "{High Velocity Outflow in the Extended Emission-Line Region of the Seyfert Galaxy NGC 7319}",
      journal = {\aj},
         year = 1996,
        month = jan,
       volume = {111},
        pages = {140},
          doi = {10.1086/117767}
}

@ARTICLE{Aromaletal25,
       author = {{Aromal}, P. and {Gallagher}, S.~C. and {Fedotov}, K. and {Bastian}, N. and {Lisenfeld}, U. and {Charlton}, J.~C. and {Appleton}, P.~N. and {Braine}, J. and {Johnson}, K.~E. and {Tzanavaris}, P. and {Emonts}, B.~H.~C. and {Togi}, A. and {Xu}, C.~K. and {Guillard}, P. and {Barcos-Mu{\~n}oz}, L. and {Smith}, L.~J. and {Konstantopoulos}, I.~S.},
        title = "{Characterizing the Star Cluster Populations in Stephan's Quintet Using HST and JWST Observations}",
      journal = {\apj},
         year = 2025,
        month = nov,
       volume = {994},
       number = {1},
          eid = {90},
        pages = {90},
          doi = {10.3847/1538-4357/ae0cb4}}

@ARTICLE{PereiraSantaellaetal22,
       author = {{Pereira-Santaella}, M. and {{\'A}lvarez-M{\'a}rquez}, J. and {Garc{\'\i}a-Bernete}, I. and {Labiano}, A. and {Colina}, L. and {Alonso-Herrero}, A. and {Bellocchi}, E. and {Garc{\'\i}a-Burillo}, S. and {H{\"o}nig}, S.~F. and {Ramos Almeida}, C. and {Rosario}, D.},
        title = "{Low-power jet-interstellar medium interaction in NGC 7319 revealed by JWST/MIRI MRS}",
      journal = {\aap},
         year = 2022,
        month = sep,
       volume = {665},
          eid = {L11},
        pages = {L11},
          doi = {10.1051/0004-6361/202244725}
}

@ARTICLE{Haynesetal18,
       author = {{Haynes}, Martha P. and {Giovanelli}, Riccardo and {Kent}, Brian R. and {Adams}, Elizabeth A.~K. and {Balonek}, Thomas J. and {Craig}, David W. and {Fertig}, Derek and {Finn}, Rose and {Giovanardi}, Carlo and {Hallenbeck}, Gregory and {Hess}, Kelley M. and {Hoffman}, G. Lyle and {Huang}, Shan and {Jones}, Michael G. and {Koopmann}, Rebecca A. and {Kornreich}, David A. and {Leisman}, Lukas and {Miller}, Jeffrey and {Moorman}, Crystal and {O'Connor}, Jessica and {O'Donoghue}, Aileen and {Papastergis}, Emmanouil and {Troischt}, Parker and {Stark}, David and {Xiao}, Li},
        title = "{The Arecibo Legacy Fast ALFA Survey: The ALFALFA Extragalactic H I Source Catalog}",
      journal = {\apj},
         year = 2018,
        month = jul,
       volume = {861},
       number = {1},
          eid = {49},
        pages = {49},
          doi = {10.3847/1538-4357/aac956}
}

@ARTICLE{Springobetal05,
       author = {{Springob}, Christopher M. and {Haynes}, Martha P. and {Giovanelli}, Riccardo and {Kent}, Brian R.},
        title = "{A Digital Archive of H I 21 Centimeter Line Spectra of Optically Targeted Galaxies}",
      journal = {\apjs},
         year = 2005,
        month = sep,
       volume = {160},
       number = {1},
        pages = {149-162},
          doi = {10.1086/431550}
}

@ARTICLE{Leroy2008,
       author = {{Leroy}, Adam K. and {Walter}, Fabian and {Brinks}, Elias and {Bigiel}, Frank and {de Blok}, W.~J.~G. and {Madore}, Barry and {Thornley}, M.~D.},
        title = "{The Star Formation Efficiency in Nearby Galaxies: Measuring Where Gas Forms Stars Effectively}",
      journal = {\aj},
         year = 2008,
        month = dec,
       volume = {136},
       number = {6},
        pages = {2782-2845},
          doi = {10.1088/0004-6256/136/6/2782}
          }

@ARTICLE{Schruba2011,
       author = {{Schruba}, Andreas and {Leroy}, Adam K. and {Walter}, Fabian and {Bigiel}, Frank and {Brinks}, Elias and {de Blok}, W.~J.~G. and {Dumas}, Gaelle and {Kramer}, Carsten and {Rosolowsky}, Erik and {Sandstrom}, Karin and {Schuster}, Karl and {Usero}, Antonio and {Weiss}, Axel and {Wiesemeyer}, Helmut},
        title = "{A Molecular Star Formation Law in the Atomic-gas-dominated Regime in Nearby Galaxies}",
      journal = {\aj},
         year = 2011,
        month = aug,
       volume = {142},
       number = {2},
          eid = {37},
        pages = {37},
          doi = {10.1088/0004-6256/142/2/37}
}

@ARTICLE{Roberts2021,
       author = {{Roberts}, I.~D. and {van Weeren}, R.~J. and {McGee}, S.~L. and {Botteon}, A. and {Ignesti}, A. and {Rottgering}, H.~J.~A.},
        title = "{LoTSS jellyfish galaxies. II. Ram pressure stripping in groups versus clusters}",
      journal = {\aap},
         year = 2021,
        month = aug,
       volume = {652},
          eid = {A153},
        pages = {A153},
          doi = {10.1051/0004-6361/202141118}
}

\appendix
\restartappendixnumbering

\renewcommand{\thetable}{A\arabic{table}}
\setcounter{table}{0}

\startlongtable
\begin{deluxetable*}{lccccccc}
\tablecaption{Neighboring Galaxies Detected with MeerKAT in H\,{\sc i}
\label{tab:other_galaxies}}
\tabletypesize{\footnotesize}
\tablehead{
\colhead{ID} &
\colhead{R.A.} &
\colhead{Decl.} &
\colhead{H\,{\sc i} Velocity} &
\colhead{Flux density} &
\colhead{$M_{\mathrm{H\,I}}$} &
\colhead{Optical/IR ID} & 
\colhead{References} \\
\colhead{} &
\colhead{(J2000)} &
\colhead{(J2000)} &
\colhead{(km s$^{-1}$)} &
\colhead{$(\rm Jy\, km s^{-1})$} &
\colhead{$(\times10^{9}M_\odot)$} &   
\colhead{} & 
\colhead{}
}
\startdata
Anon 1    & 22:35:41.3 & +34:16:01 & 6254 & $0.50\pm0.05$  & $1.02 \pm 1.00$ & WISEA J223541.84+341601.6 & 1,4 \\ 
Anon 2    & 22:35:50.4 & +34:09:30 & $6377$ & $0.38\pm0.04$  & $0.78 \pm 0.08$ & SDSS J223550.70+340930.8 & 1,2,3,4 \\ 
Anon 4    & 22:36:40.4 & +33:56:56 & $6064$ & $0.93\pm0.10$  & $1.90 \pm 0.40$ & SDSS J223640.85+335654.5 & 1,2,3,4 \\ 
Anon 6  & 22:37:13.8 & +34:05:06 & 6072 & $0.82\pm0.10$ & $1.67 \pm 0.20$ & WISEA J223712.99+340508.9 & 2, 4\\ 
Anon 8  & 22:35:47.5 & +34:06:48 & $5724$ & $0.67\pm0.06$  & $0.14 \pm 0.01$ & SDSS J223547.54+340648.1 & 3 \\ 
NGC 7320C & 22:36:20.7 & +33:59:09 & 6520 & $0.098\pm0.01$& $0.20 \pm 0.02$ & NGC 7320C & - \\ 
S1  & 22:37:26.9 & +33:55:25 & 6359 & $0.78\pm0.08$ & $1.60\pm0.20$   & NGC 7320B & 4 \\ 
S2  & 22:36:47.7 & +33:41:26 & 6818 & $0.72\pm0.08$ & $1.47\pm0.02$ & WISEA J223648.42+334131.3 & - \\ 
S3  & 22:38:49.9 & +34:18:32 & 6267 & $0.053\pm0.005$ & $0.11\pm0.01$ & UGC 12132 & 1,4,5 \\ 
S4  & 22:37:25.8 & +34:22:02 & 6628 & $0.38\pm0.04$ & $0.78\pm0.08$ & NGC 7337 & 4 \\ 
S5 & 22:34:36.2 & +33:40:06 & 6300 & $0.31\pm0.03$ &$0.63\pm0.06$ & WISEA J223437.66+334006.3 & - \\
S6 & 22:34:31.1 & +33:52:38 & 5966 & $0.39\pm0.04$ & $0.78\pm0.08$ & WISEA J223431.13+335250.2 & - \\
S7  & 22:36:59.4 & +33:49:56 & $7278$ & $0.13\pm0.01$ & $0.26 \pm 0.02$  & WISEA J223659.36+334948.7 & - \\ 
S8 & 22:35:26.3& +34:08:37 & 6075  & $0.48\pm0.05$  & $0.98\pm0.10$ & SDSS J223526.26+340837.9 & - \\
S9 & 22:38:40.0 & +34:20:02 & 6332 & $0.31\pm0.03$ & $0.63\pm0.06$ & WISEA J223840.57+342016.3 & - \\
S10  & 22:38:34.3 & +34:04:12 & 7450 & $2.25\pm0.20$ & $4.59\pm0.41$ & NGC~7343 & 5 \\ 
\enddata
\tablerefs{prior \Hi\ detections: 1: \citet{Shostak1984} 2: \citet{Williams2002}, 3: \citet{Xu2022,Cheng2023}, 4: \citet{Haynesetal18}, 5: \citet{Springobetal05}}
\end{deluxetable*}

\renewcommand{\thefigure}{A.\arabic{figure}}


\begin{figure*}[!thbp]
    \centering
    \includegraphics[width=0.44\textwidth]{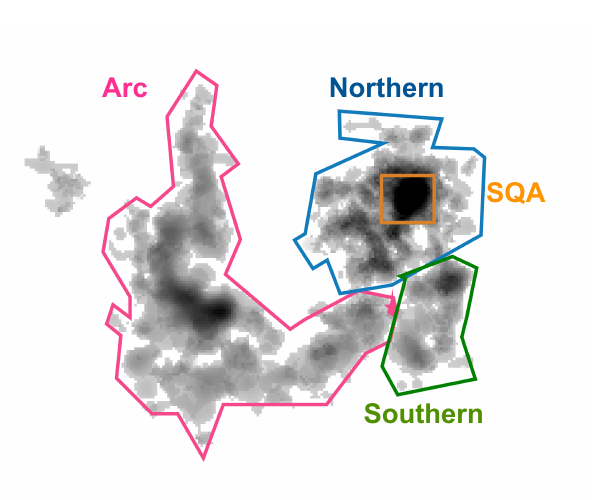}
        \includegraphics[width=0.44\textwidth]{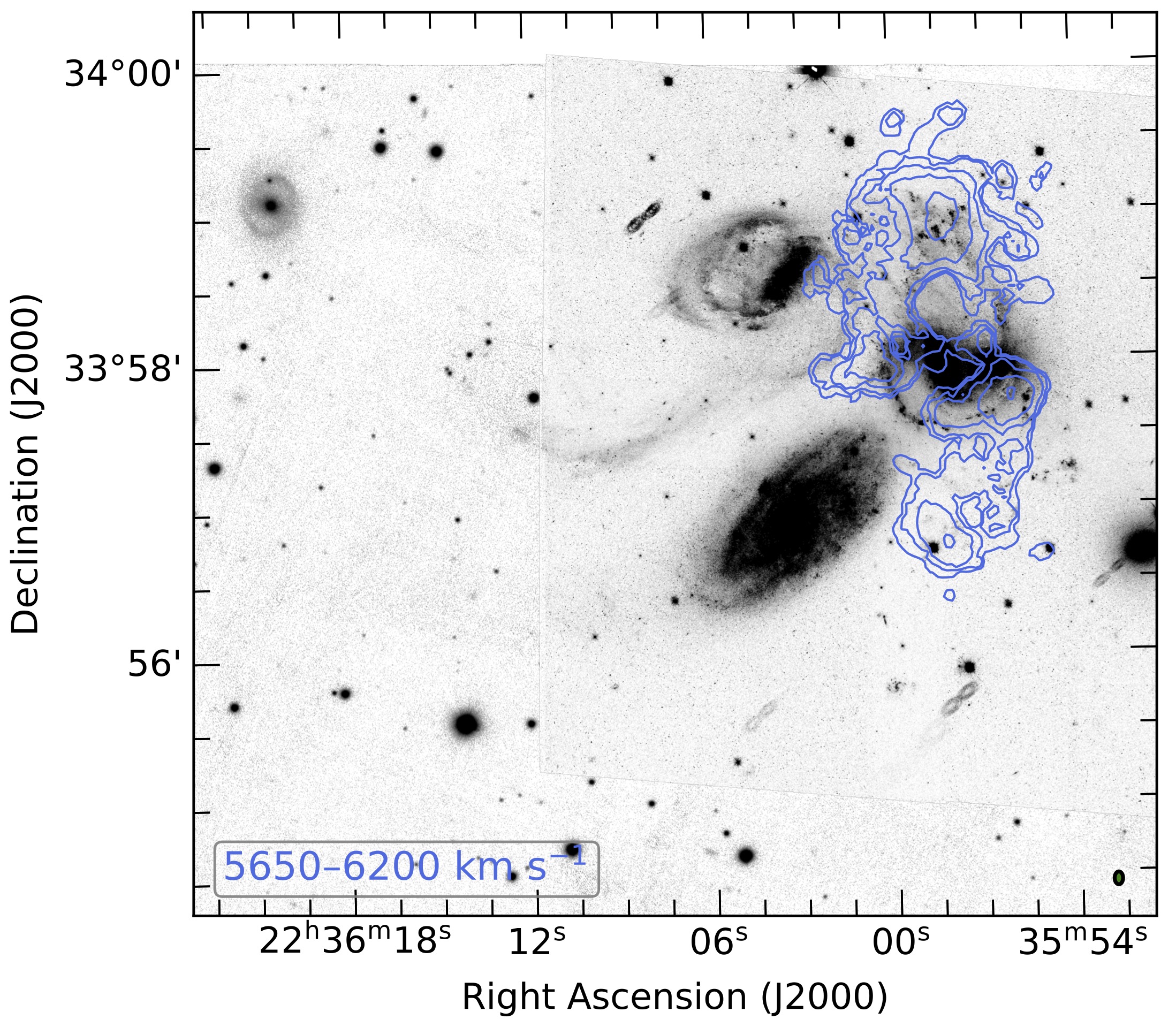}
    \vspace{-0.2cm}
 \caption{\textit{Left}: MeerKAT medium resolution ($17\arcsec\times14\arcsec$) total intensity \hi~ map superimposed with regions used to extract the spectra shown in Figure\,\ref{fig::HI_Spectra}. To measure the total integrated flux density of Arc, Northern and Southern regions we used the low resolution map (i.e., 25\arcsec\ beam size). \textit{Right}: MeerKAT moment-zero map of the combined 5700 and 6000\kmps\ components. \Hi\ contour levels are plotted at $5.6, 11.1, 22.3, 44.5, 89.1, 178 ...... \times 10^{19}\rm \,atoms\,cm^{-2}$. The radio beam size is $17\arcsec\times14\arcsec$.}
      \label{fig::HI_regions}
\end{figure*} 

\begin{figure*}[!thbp]
    \centering
    \includegraphics[width=0.7\textwidth]{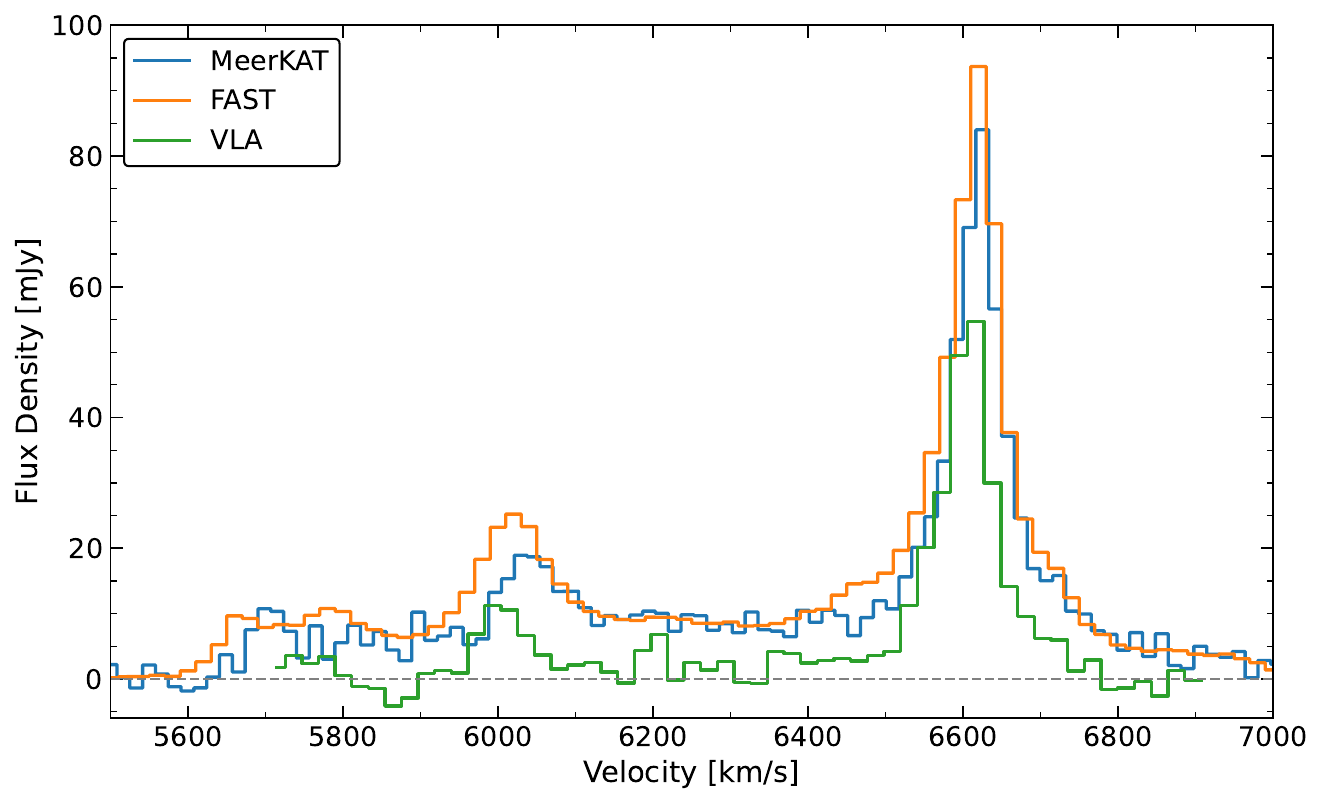}
    \vspace{-0.2cm}
 \caption{Comparison of MeerKAT, FAST \citep{Cheng2023} and VLA \citep{Jones2023} integrated \hi~spectra for \SQ. The overall \Hi\ profile shape  for obtained from MeerKAT is similar to that measured by FAST. The dashed gray horizontal line indicates the zero flux density level.}
      \label{fig::FAST_MeerKAT}
\end{figure*}

\begin{figure*}[!thbp]
    \centering
    \includegraphics[width=0.49\textwidth]{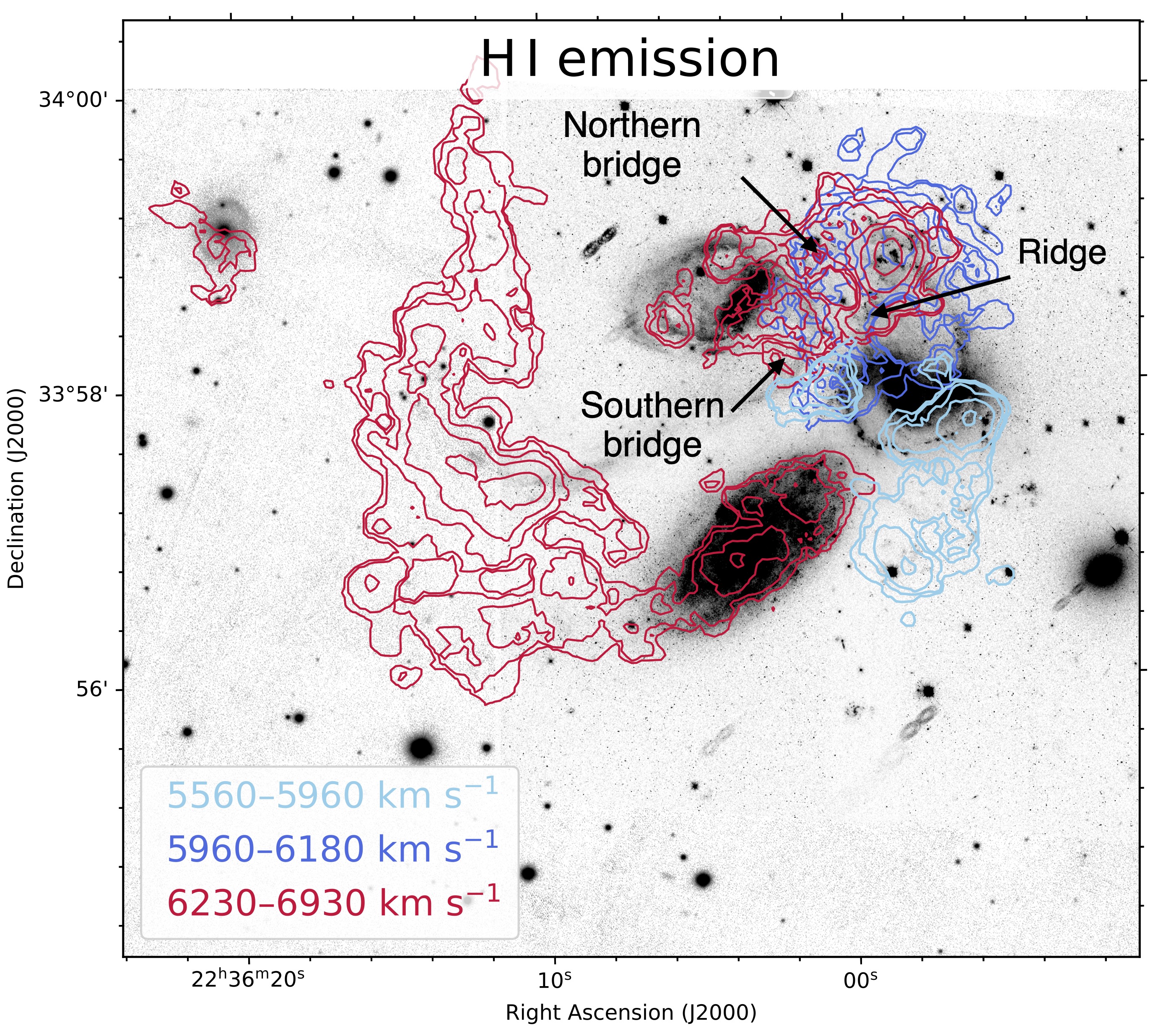}
    \includegraphics[width=0.49\textwidth]{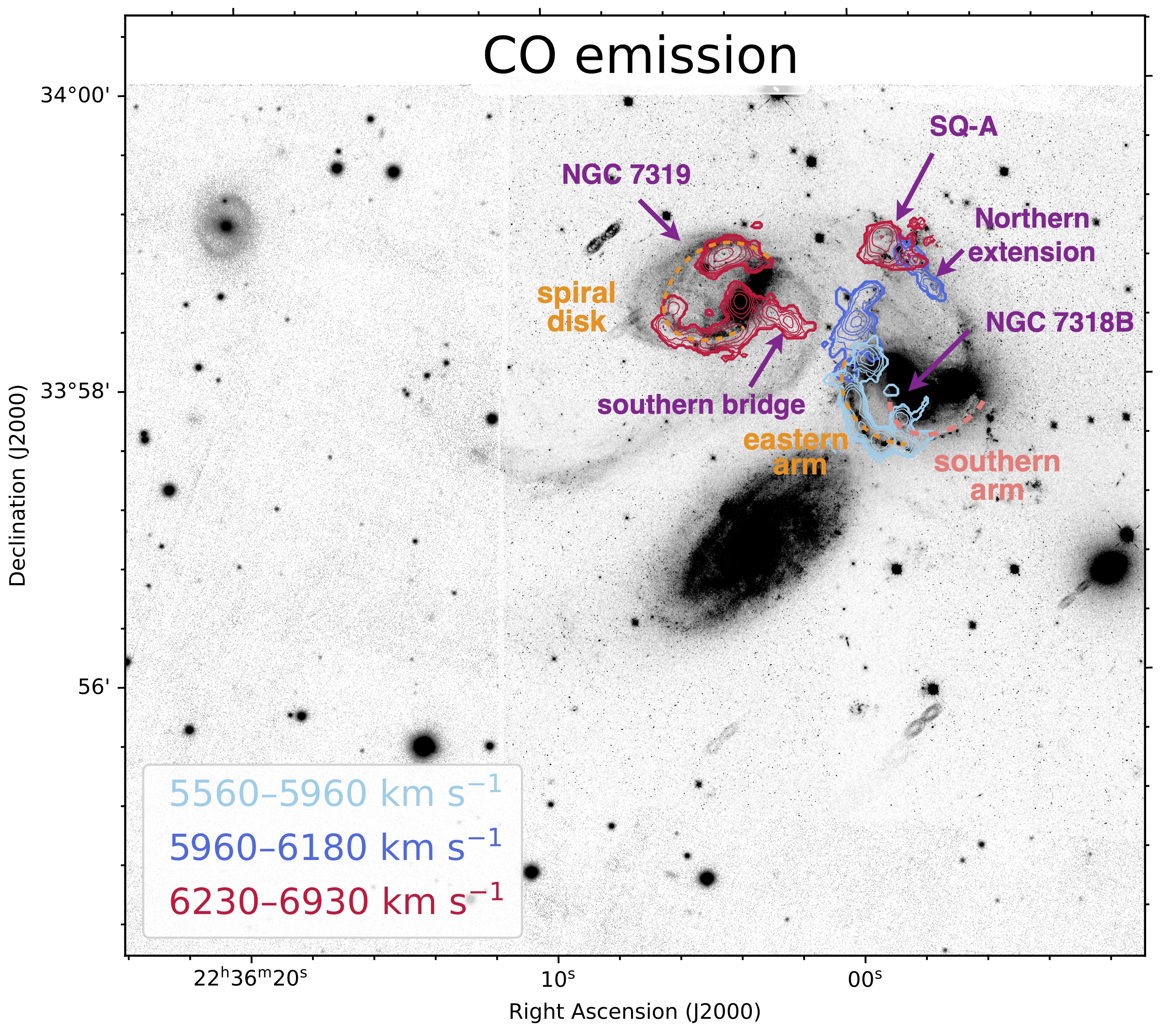}
 \caption{ Kinematic structures of the atomic and cold molecular gases in \SQ. The MeerKAT \Hi\ (left) and CO (right) emission are superimposed on the HST+DSS optical image, integrated across three velocity range. The \Hi\ contours are same as in Figure\,\ref{fig::SQ_HI_labeling}. The CO(2-1) contours are from Atacama Compact Array with the beam size $8\arcsec\times7\arcsec$ \citep[adopted from][]{Emonts2025}. The CO contours start at $8\,\rm mJy\,beam^{-1}$ and increase by a factor 2. }
      \label{fig::H1_CO}
\end{figure*}

\begin{figure*}[ht]
\centering
\includegraphics[width=1.0\textwidth]{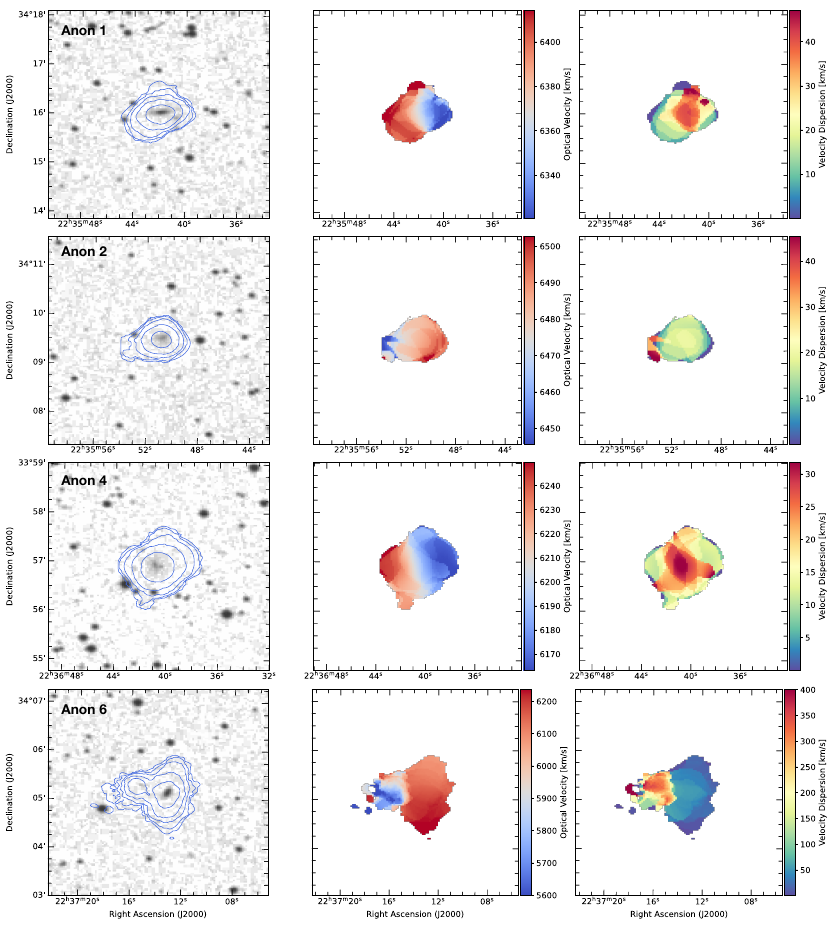}
\caption{MeerKAT \hi~ column density distribution contours  overlaid on DSS (left),  first-order moment of the radial velocity field (middle) and and the second-order moment of the velocity dispersion of neighboring galaxies in the field.  \Hi\ contour levels are plotted at $5.6, 11.1, 22.3, 44.5, 89.1, 178 ...... \times 10^{19}\rm \,atoms\,cm^{-2}$. The beam size is $25\arcsec\times 25\arcsec$.}
\label{fig:appendix1}
\end{figure*}

\begin{figure*}[ht]
\centering
\includegraphics[width=1.0\textwidth]{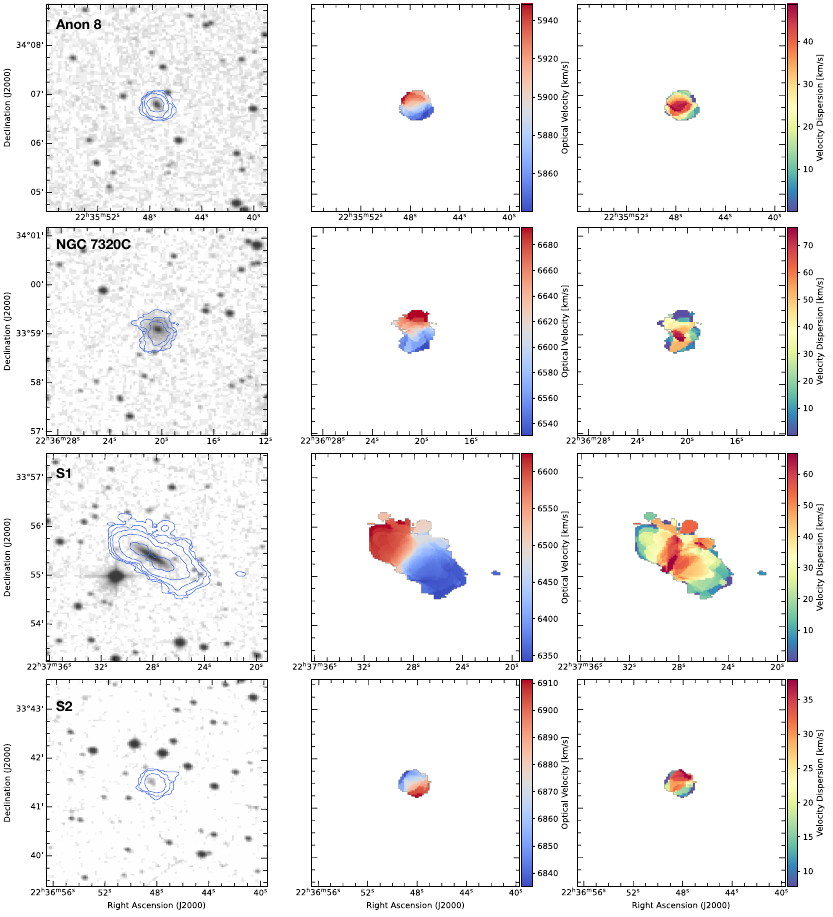}
\caption{Continued. The beam size and \hi~contour levels are the same as in Figure\,\ref{fig:appendix1}.}
\label{fig:appendix2}
\end{figure*}

\begin{figure*}[ht]
\centering
\includegraphics[width=1.0\textwidth]{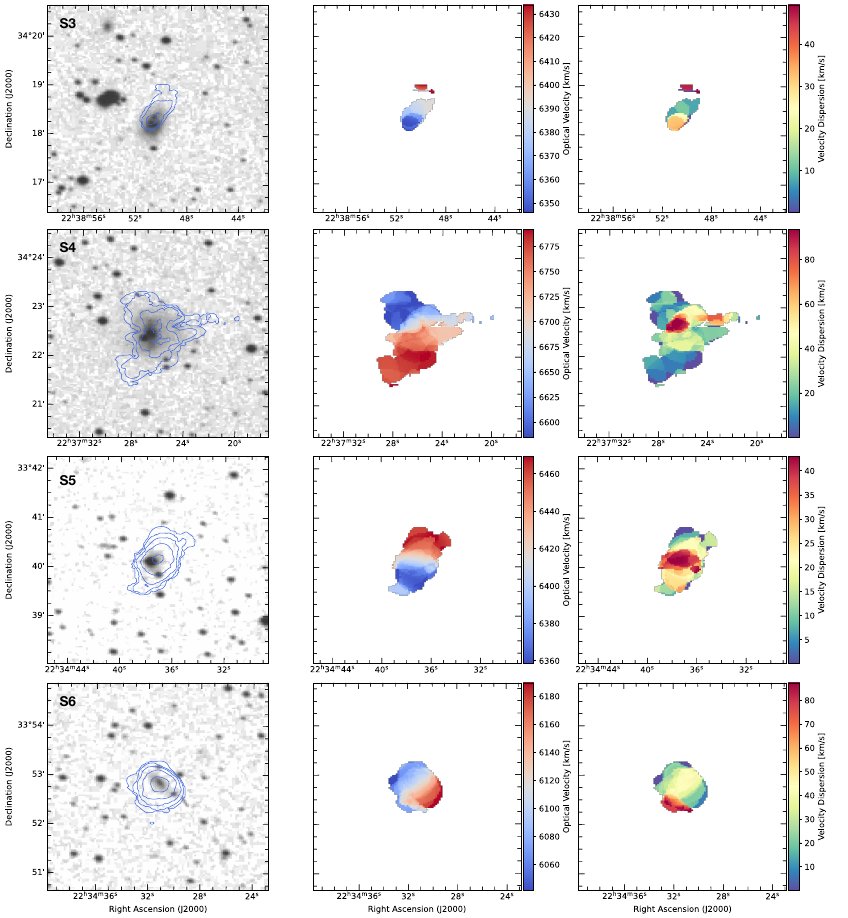}
\caption{Continued. The beam size and \hi~contour levels are the same as in Figure\,\ref{fig:appendix1}.}
\label{fig:appendix3}
\end{figure*}

\begin{figure*}[ht]
\centering
\includegraphics[width=1.0\textwidth]{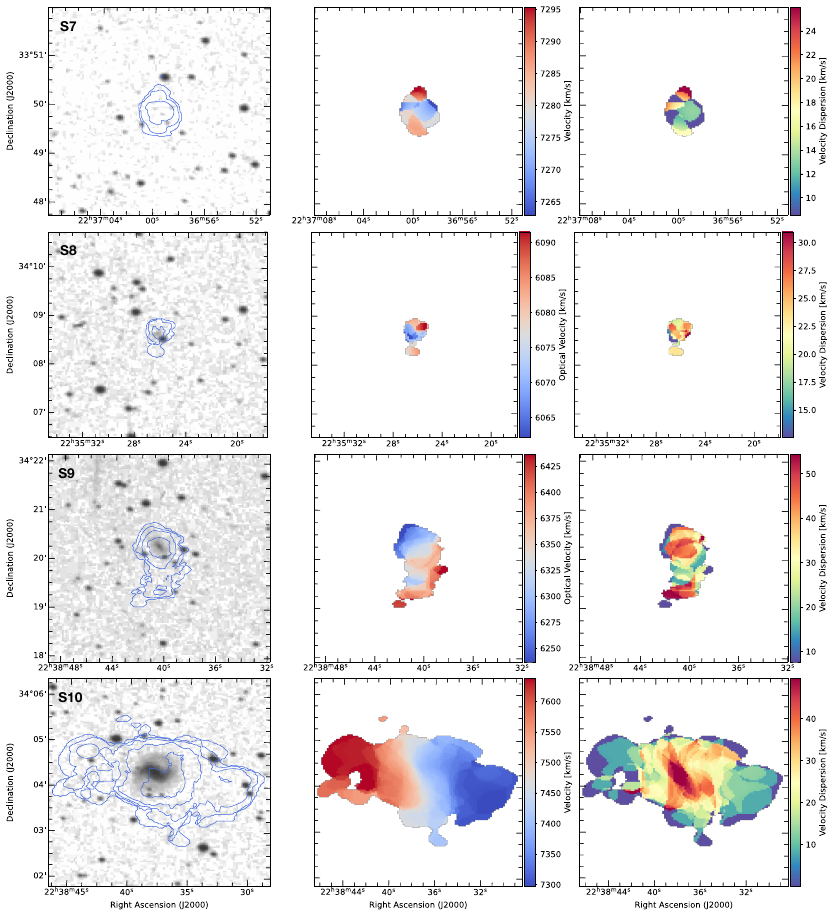}
\caption{Continued. The beam size and \hi~contour levels are the same as in Figure\,\ref{fig:appendix1}.}
\label{fig:appendix4}
\end{figure*}

\end{document}